\documentclass[lettersize,journal]{IEEEtran}
\usepackage{amsmath,amsfonts}
\usepackage{algorithmic}
\usepackage{algorithm}
\usepackage{array}
\usepackage[caption=false,font=normalsize,labelfont=sf,textfont=sf]{subfig}
\usepackage{textcomp}
\usepackage{stfloats}
\usepackage{url}
\usepackage{verbatim}
\usepackage{graphicx}
\usepackage{cite}
\usepackage{comment}

\usepackage{booktabs}

\usepackage{adjustbox}
\usepackage{enumitem}
\usepackage{hyperref}
\begin{document}
\title{Audio-visual video-to-speech synthesis with synthesized input audio}

\author{Triantafyllos Kefalas, Yannis Panagakis,~\IEEEmembership{Member,~IEEE}, Maja Pantic,~\IEEEmembership{Fellow,~IEEE}
\thanks{This work has been submitted to the IEEE for possible publication. Copyright may be transferred without notice, after which this version may no longer be accessible. Corresponding author: Triantafyllos Kefalas (email: tk15@imperial.ac.uk)}
\thanks{Triantafyllos Kefalas and Maja Pantic are with the Department of Computing, Imperial College London, UK}
\thanks{Yannis Panagakis is with the Department of Informatics and Telecommunications, University of Athens, Greece}}

\markboth{}%
{Shell \MakeLowercase{\textit{et al.}}: A Sample Article Using IEEEtran.cls for IEEE Journals}


\maketitle

\begin{abstract}
Video-to-speech synthesis involves reconstructing the speech signal of a speaker from a silent video. The implicit assumption of this task is that the sound signal is either missing or contains a high amount of noise/corruption such that it is not useful for processing. Previous works in the literature either use video inputs only or employ both video and audio inputs during training, and discard the input audio pathway during inference. In this work we investigate the effect of using video and audio inputs for video-to-speech synthesis during both training and inference. In particular, we use pre-trained video-to-speech models to synthesize the missing speech signals and then train an audio-visual-to-speech synthesis model, using both the silent video and the synthesized speech as inputs, to predict the final reconstructed speech. Our experiments demonstrate that this approach is successful with both raw waveforms and mel spectrograms as target outputs.
\end{abstract}

\begin{IEEEkeywords}
Video-to-speech, speech synthesis, generative adversarial networks (GANs), conformer, audio-visual
\end{IEEEkeywords}

\section{Introduction}
\IEEEPARstart{D}{eep} learning has revolutionized the field of automatic speech recognition (ASR), paving the way for communication between humans and machines across multiple domains such as autonomous vehicles\cite{driven_to_distraction}, virtual assistants\cite{manifestation_of_virtual_assistants} and voicebots for services including digital banking\cite{intelligent_voice_bots_for_digital_banking}. However, the performance of ASR systems degrades in the presence of acoustic noise \cite{deep_av_speech_recognition}. This has motivated research that leverages the multi-modal nature of speech, and in particular the co-occurrence of spoken words and lip movements, to conduct ASR with both visual and audio inputs \cite{deep_av_speech_recognition, end_to_end_av_speech_recognition, end_to_end_av_speech_recognition_with_conformers, av_speech_recognition_using_bimodal_trained_bottleneck_features, av_speech_recognition_with_hybrid_ctc_attention_architecture}. Lipreading, i.e., predicting text from a silent video, has also been investigated in situations where the audio modality is missing or very noisy \cite{deep_learning_based_automated_lipreading_a_survey, review_on_research_progress_of_machine_lipreading}.

Although lipreading has undergone significant progress in recent years, synthesizing the missing (or noisy) audio from the video modality is beneficial in multiple scenarios. Firstly, this is an additional and indirect approach to perform lipreading, i.e., one may synthesize the missing audio from the silent video first and then perform ASR on the synthesized audio. For example, several video-to-speech works employ pre-trained ASR models to compute the word error rate of their synthesized audio samples (e.g., \cite{video_driven_speech_reconstruction_using_gans, end_to_end_video_to_speech_synthesis_using_gans, lip_to_speech_synthesis_with_visual_context_attention_gan, svts}). Secondly, there are several applications that make use of the synthesized audio, including speech enhancement for videoconferecing \cite{ephrat2017vid2speech}, speech reconstruction in silent videos from surveillance cameras\cite{ephrat2017vid2speech, av_event_recognition_in_surveillance}, synthesizing speech for individuals with aphonia \cite{end_to_end_video_to_speech_synthesis_using_gans}, and making silent speech interfaces for environments requiring silence \cite{silent_speech_interfaces}.

The above considerations have led to increasing interest in the task of video-to-speech synthesis (V2A), involving the reconstruction of the speech signal from a silent video. Although one possible approach is to predict text with a lipreading model first followed by audio synthesis with a text-to-speech (TTS) system, this entails multiple disadvantages: one requires text labels, the text modality does not contain information about a speaker's voice, and the performance of the lipreading model bounds the accuracy of the synthesized speech. Thus, several V2A methods have been introduced that predict either intermediate acoustic features (including linear predictive codes\cite{ephrat2017vid2speech} and mel spectrograms \cite{lip2wav, lip_to_speech_synthesis_with_visual_context_attention_gan, speech_reconstruction_visual_voice_memory, lipsound2}) followed by vocoder synthesis, or raw waveforms directly \cite{video_driven_speech_reconstruction_using_gans, end_to_end_video_to_speech_synthesis_using_gans}. Most of these works train models with silent video inputs only, although some employ audio inputs as well \cite{lipsound2, speech_reconstruction_visual_voice_memory}. For example, \cite{lipsound2} employs teacher forcing during training using the ground truth mel spectrograms as inputs, and \cite{speech_reconstruction_visual_voice_memory} learns face and voice associations using a key-value memory structure that discards the audio encoder after training. However, none of these methods have investigated the possibility of including audio inputs (in addition to the silent video) during inference as well. 

\begin{figure*}[t]
  \includegraphics[width=\linewidth]{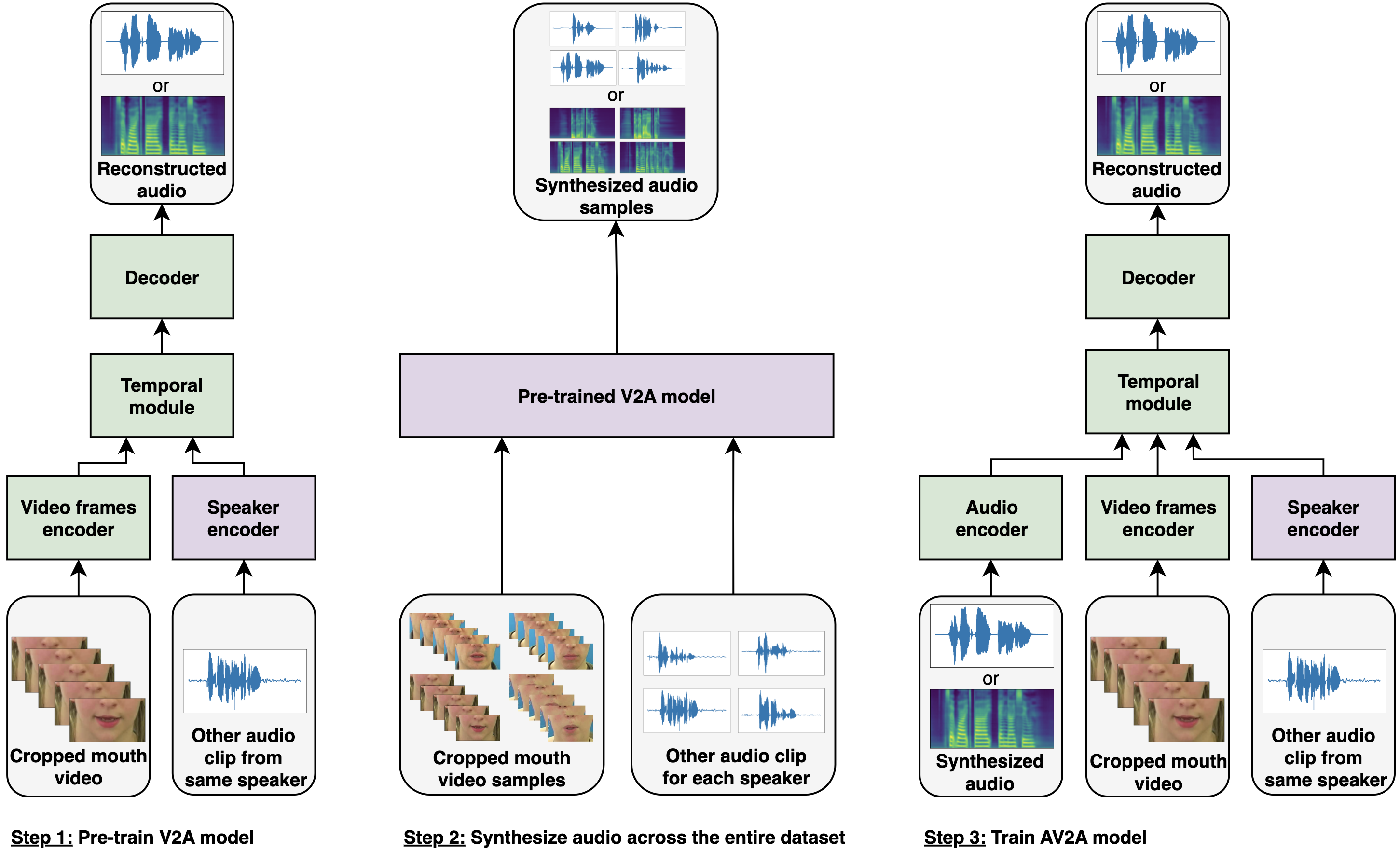}
\caption{High-level overview of the proposed method. Green modules are trainable, purple modules are pre-trained and frozen.}
\label{fig:high_level_av2a}
\end{figure*}

In this work we investigate video-to-speech synthesis models, following an encoder-decoder structure, that include audio and video inputs during both training and inference. We do so in both the raw waveform and mel spectrogram domains, as illustrated in Fig. \ref{fig:high_level_av2a}. Given that the V2A task posits we do not have access to the audio samples of input videos  during inference, we first synthesize them using a pre-trained V2A model. Once we have obtained the synthesized audio samples we combine them with their corresponding silent video samples as inputs to an audio-visual-to-speech model (AV2A). To obtain the latter, we propose appending an audio encoder to the V2A model used above, thus constructing a corresponding AV2A model. This approach, which involves transforming a V2A model to an AV2A model and using the V2A model's synthesized samples, has the additional benefit of allowing us to investigate whether a given V2A model can be improved by using its own outputs as self-supervision. Hence, we refer to the V2A model in question as the base model to highlight that we use it both to synthesize the input audio, and as the starting point to construct the AV2A model.

In addition we introduce three procedures for training the AV2A models. The first, baseline approach, involves reconstructing audio directly from the audio-visual inputs. Inspired by recent work in audio-visual self-supervised learning\cite{av_hubert}, the second procedure employs modality dropout, whereby we train in an alternating fashion between the baseline approach above, dropping the audio modality and dropping the visual modality. Finally, we also train with the aforementioned modality dropout, modified such that we use the ground truth audio, instead of the synthesized audio, when dropping the visual modality (i.e., when using the audio modality only). Modality dropout is also motivated by the observation that conditional models ignore parts of their input when the conditioning information is very strong (e.g. ignoring the noise vector in conditional image GANs \cite{deep_multi_scale_video_prediction_beyond_mse, image_to_image_translation_with_cgans}, and the presence of the audio modality dominating predictions in audio-visual speech recognition \cite{av_hubert, deep_av_speech_recognition, end_to_end_av_speech_recognition_with_conformers}). By dropping a modality, our models are forced to reconstruct the audio using the remaining modality only, thus avoiding the trivial solution of relying on the ``easier'' modality for reconstruction.  

Our contributions are summarized as follows:
\begin{itemize}
\item We propose a video-to-speech synthesis approach that employs audio and video inputs during both training and inference by transforming the V2A model into a corresponding AV2A model, and using the former to synthesize the audio inputs. We do so in both the raw waveform and the mel spectrogram domains.

\item We introduce two versions of training with modality dropout \cite{av_hubert}, which involves training by alternating between the input modalities (audio only, visual only, audio-visual).

\item We modify the batch normalization \cite{batch_norm} layers in our models such that they keep track of separate running statistics for the audio, video and audio-visual modalities (i.e., for each setting of modality dropout).

\item We carry out experiments on popular audio-visual face and speech datasets on seen (GRID \cite{grid_database}, TCD-TIMIT \cite{tcd_timit_databse}) and unseen (GRID\cite{grid_database}, LRW\cite{lip_reading_in_the_wild}) speakers.
\end{itemize}

\section{Related work}
\subsection{Lipreading}
Lipreading, also known as visual speech recognition, involves predicting a speaker's words from silent videos. It has been a longstanding research interest in the speech processing community and was originally proposed in \cite{visual_contribution_to_speech_intelligibility_in_noise}, particularly for supplementing audio speech recognition in environments with high acoustic noise. The feasibility of computer-based lipreading was demonstrated in early experiments in both visual and audio-visual speech recognition in \cite{continuous_optical_automatic_speech_recognition_by_lipreading} and \cite{automatic_lipreading_to_enhance_speech_recognition} respectively.

Prior to the widespread adoption of deep learning, lipreading methods employed handcrafted visual features used as input to Hidden Markov Models (HMMs) \cite{information_theoretic_feature_extraction_for_av_speech_recognition} or Support Vector Machines (SVMs) \cite{lipreading_with_local_spatiotemporal_descriptors} to predict text. These include DCT features \cite{information_theoretic_feature_extraction_for_av_speech_recognition}, DWT features \cite{an_image_transform_approach_for_hmm_based_automatic_lipreading}, geometric features \cite{lip_feature_extraction_and_reduction_for_hmm, lip_feature_extraction_based_on_improved_jumping_snake_model, profile_view_lip_reading} and Active Appearance Models \cite{active_appearance_models, view_independent_computer_lip_reading, insights_into_machine_lip_reading}. In recent years, research has focused on end-to-end deep learning (DL) models which have outperformed traditional methods. Early DL lipreading works studied word and phoneme-level prediction in two stages: extracting deep features first, then training a classifier \cite{lipreading_using_convolutional_neural_network, deep_learning_of_mouth_shapes_for_sign_language, improved_speaker_independent_lip_reading, av_speech_recognition_using_bimodal_trained_bottleneck_features, deep_complementary_bottleneck_features_for_visual_speech_recognition}. Lipnet \cite{lipnet} was the first end-to-end work on sentence-level prediction and comprised a spatio-temporal convolutional encoder, a bidirectional GRU \cite{empirical_evaluation_of_gated_rnn_on_sequence_modeling} and a Linear layer. Trained with a CTC loss \cite{connectionist_temporal_classification}, it produced state-of-the-art results on the GRID database \cite{grid_database}.

Other end-to-end lipreading models have focused on predictions at the word and character level. For example, in \cite{end_to_end_visual_speech_recognition_with_lstms, end_to_end_visual_speech_recognition_for_small_scale_datasets} a model with fully-connected layers and LSTMs predicted words from cropped mouth regions and their corresponding difference images. A CNN-based model was proposed in \cite{lip_reading_in_the_wild} and applied to silent videos of single-word utterances, recorded in-the-wild. In \cite{large_scale_visual_speech_recognition} a CNN+LSTM based model that predicted phonemes was presented, followed by word decoding, while \cite{combining_residual_networks_with_lstms_for_lipreading} introduced a word-level prediction model with residual connections and LSTMs.

By incorporating the audio modality as well (i.e. performing audio-visual speech recognition) one may leverage the natural co-occurrence of audio and video to obtain superior results to a unimodal model, for example by distinguishing homophenes. In \cite{end_to_end_av_speech_recognition} two ResNet-based encoders, followed by bidirectional GRUs, encode the mouth frames and raw audio respectively, while in \cite{deep_av_speech_recognition} ResNet+Transformer back-end is employed. For further coverage of lipreading methods and approaches, we refer the reader to relevant surveys \cite{deep_learning_based_automated_lipreading_a_survey, review_on_research_progress_of_machine_lipreading}.
\subsection{Video-to-speech synthesis}

Early work in video-to-speech synthesis\cite{reconstructing_intelligible_audio_speech_from_visual_speech_features, generating_intelligible_audio_speech_from_visual_speech} relied on handcrafted visual features, such as 2D-DCT and AAM features. To the best of our knowledge, \cite{reconstructing_intelligible_audio_speech_from_visual_speech_features} was the first work predicting speech directly from visual inputs and involved a model receiving visual features and using Gaussian Mixture Models (GMMs) or fully-connected neural networks to to predict spectral envelope representations (LPCs or mel-filterbank amplitudes). A STRAIGHT vocoder \cite{straight_vocoder} then processed the output and synthesized raw waveforms. In \cite{generating_intelligible_audio_speech_from_visual_speech} this was extended, where a classification framework was employed to predict codebook entries corresponding to audio vectors, resulting in better speech intelligibility.

A model with CNNs was employed in \cite{ephrat2017vid2speech} that learned features from raw pixels (grayscale video frames), outputting line spectral pairs (LSP) features. 
These were fed with Gaussian noise to a source-filter speech synthesizer to generate the raw waveforms. In \cite{improved_speech_reconstruction_from_silent_video} this was improved by
employing one ResNet encoder for raw video and another one for optical flow. The two sets of embeddings are then concatenated and fed into a fully-connected network to output mel spectrograms, followed by post-processing module to generate linear spectrograms and the Griffin-Lim algorithm (GLA) \cite{griffin_lim} to obtain the raw waveforms. Lip2AudSpec \cite{lip2audspec} first trains a fully-connected auto-encoder on spectrograms and then uses the learned bottleneck features as training targets for a CNN+RNN lipreading model. A multi-task model was presented in \cite{vocoder_based_speech_synthesis}, which predicts the spectral envelope, aperiodic parameters and the fundamental frequency as inputs to a vocoder to synthesize the raw waveform. It is also trained jointly for a lipreading task using a connectionist temporal classification (CTC) \cite{connectionist_temporal_classification} loss. Lip2Wav \cite{lip2wav} uses an encoder-decoder architecture to generate mel spectrograms from video frames. The model features a stack of 3D convolutions and an attention-based decoder inspired by Tacotron2\cite{tacotron2}, with GLA applied to extract the raw waveforms from the mel spectrograms.

In VCA-GAN \cite{lip_to_speech_synthesis_with_visual_context_attention_gan} the input video sequence is summarized and included as a global conditioning variable and a multi-scale generator with residual blocks is used to generate mel spectrograms from coarse to fine-level. This is followed by a post-net, similar to that proposed in \cite{improved_speech_reconstruction_from_silent_video}, and GLA to synthesize the raw waveforms. Along with a multi-scale discriminator, it is trained with an adversarial loss, reconstruction and synchronization losses.

In addition, a VAE-based\cite{vae_paper} model was introduced in \cite{speech_prediction_in_silent_videos_using_vaes}, in order to model the uncertainty in generating speech. This involves an encoder-decoder model, where the encoder and decoder are bridged by LSTMs to model the uncertainty autoregressively, using conditional probability distributions. The model generates mel spectrograms, and GLA is applied to obtain the raw waveforms.

Visual Voice Memory \cite{speech_reconstruction_visual_voice_memory} proposes a key-value memory structure, inspired by memory networks \cite{key_value_memory_networks}, to map visual features (keys) to audio features (values). During training, the model employs video and speech encoders to extract features from the video frames and ground truth mel spectrograms respectively. The decoder synthesizes mel spectrograms by first concatenating the visual features with these audio features, or with audio features as predicted by the intermediate memory module. The audio encoder is discarded during inference, and the memory module receives the visual features to output imprinted audio features, to be concatenated with the visual features and fed to the decoder.

LipSound2 \cite{lipsound2} proposed an encoder-decoder model with an auto-regressive bridge module in the middle. During training, teacher forcing is employed, whereby the ground truth mel spectrogram is passed as input to the bridge module, along with the visual features. During inference, the predicted outputs from previous timesteps are employed instead, in an auto-regressive manner. A highlight of this work is that it investigated the effect of pre-training this video-to-speech model, along with a speech recognition model, on VoxCeleb2 \cite{voxceleb2_paper} and LibriSpeech \cite{librispeech} respectively. The video-to-speech model was then fine-tuned on an audio-visual dataset of interest, followed by fine-tuning of the speech recognition model on the predicted audio, achieving state-of-the-art results on GRID and TCD-TIMIT.

In order to scale V2A models to large datasets in a principled manner, SVTS \cite{svts} introduced a 3D Convolution+ResNet-18 encoder followed by a conformer decoder that generates mel spectrograms and a pre-trained neural vocoder. The decoder is of varying depth, in proportion to the size of the audio-visual dataset in question. It is the first V2A model applied to LRS3 \cite{lrs3} and also achieved state-of-the-art results on the GRID and LRW \cite{lip_reading_in_the_wild} datasets.

Some recent works have employed GANs to synthesize raw waveforms from silent videos directly, trained end-to-end. In \cite{video_driven_speech_reconstruction_using_gans} an encoder-decoder GAN was proposed, whose generator is fed a silent video and outputs a raw waveform. It consists of a convolutional decoder, a GRU and a decoder with transposed convolutions. Using a convolutional waveform critic, the model is trained with a Wasserstein GAN loss\cite{wasserstein_gan, improved_training_of_wasserstein_gans} and three reconstruction losses. This work was extended in \cite{end_to_end_video_to_speech_synthesis_using_gans}, where the convolutional encoder was replaced by a 3D Convolution+ResNet-18 and a spectrogram critic was added as well.

Finally, our previous work \cite{large_scale_unsupervised_audio_pretraining_for_v2a} investigated the effect of pre-training the decoder of V2A encoder-decoder models on large volumes of audio data, followed by fine-tuning the V2A model on an audio-visual dataset of interest. This involved training audio-to-audio (A2A) encoder-decoder models on c. 3,500 hours of audio data and then initializing the decoder of a V2A model with that of a pre-trained A2A model. To this end, we proposed a family of V2A models in the raw waveform and mel spectrogram domains (inspired by the Wasserstein GAN in \cite{end_to_end_video_to_speech_synthesis_using_gans} and SVTS\cite{svts}, and denoted by V2A-WaveGAN and V2A-MelSpec respectively) and their A2A counterparts. Our experiments across GRID, TCD-TIMIT and LRW demonstrated that in most cases, this unsupervised pre-training step leads to higher quality audio, as measured by objective metrics.
\subsection{Audio-visual speech enhancement}

Speech enhancement, involving the extraction of clean speech from a noisy or corrupt speech signal, is a well-studied problem in signal processing \cite{an_overview_of_deep_learning_based_av_speech_enhancement_and_separation, supervised_speech_separation_based_on_deep_learning}. Inspired by the bimodal nature of speech perception\cite{hearing_by_eye} many research works have also investigated audio-visual speech enhancement (AVSE), whereby the visual modality, and lip movements in particular, is incorporated. 

The relevant literature is vast\cite{an_overview_of_deep_learning_based_av_speech_enhancement_and_separation} and can be categorized according to the type of input acoustic features (e.g., spectrograms, STFT, raw waveforms), input visual features (e.g. raw video pixels, AAMs, optical flow), models (e.g. auto-encoders, VAEs \cite{vae_paper}, GANs\cite{gan_paper}), audio-visual fusion methods (e.g. concatenation-based, attention-based), training targets (e.g. spectrograms, masks, raw waveforms) and objective functions (e.g. MSE, MAE, triplet loss). Research can also be grouped according to the the type of sound distortion (e.g. additive noise, reverberation, clipping) and whether it involves the presence of other speakers (leading to the problem of speech separation), both of which motivate different models and training targets \cite{an_overview_of_deep_learning_based_av_speech_enhancement_and_separation, supervised_speech_separation_based_on_deep_learning}. 

To the best of our knowledge, the first two concurrent works in AVSE were \cite{audio_visual_enhancement_of_speech_in_noise,
audio_visual_segmentation_and_the_cocktail_party_effect}. Before the advent of deep learning, subsequent methods focused on predicting time-frequency masks\cite{source_separation_of_convolutive_and_noisy_mixtures} \cite{speaker_separation_using_visually_derived_binary_masks}, as well as other training targets \cite{avss_via_hidden_markov_models, noisy_audio_feature_enhancement_using_av_speech_data, avse_with_avcdcn, av_graphical_models_for_speech_processing, video_assisted_speech_source_separation, effective_visually_derived_wiener_filtering_for_av_speech_processing}. Early deep-learning based works include \cite{seeing_through_noise}, where the video-to-speech model in \cite{improved_speech_reconstruction_from_silent_video} is used to generate a mask to filter the noisy spectrogram. In \cite{the_conversation_deep_av_speech_enhancement} a model that jointly predicts a magnitude spectrogram mask and a phase adjustment is proposed and was applied successfully to unseen speakers in-the-wild, including to LRS2 \cite{deep_av_speech_recognition} and VoxCeleb2\cite{voxceleb2_paper}. In \cite{time_domain_av_speech_separation} a model performing speech separation on raw waveforms is proposed, trained with scale-invariant source-to-noise ratio \cite{tasnet}. Other approaches include U-Net based models \cite{muse, visualvoice} as well conditional VAEs \cite{avse_using_conditional_vaes}.

Several works have investigated deep-learning based AVSE models with direct mapping, i.e., the setting where either the raw waveform or some acoustic features constitute the training target. In \cite{visual_speech_enhancement} an encoder-decoder model is proposed that is fed both the input video and the noisy mel spectrogram to output the clean mel spectrogram. A CNN-based architecture was employed in \cite{audio_visual_speech_enhancement_using_multimodal_deep_cnns} that received noisy spectrograms and cropped mouth video inputs, and reconstructed both the clean spectrograms and the cropped mouth frames. A CNN+LSTM model is introduced in \cite{av_speech_codecs_rethinking_avse} to predict mel spectrograms, followed by a pre-trained neural vocoder to reconstruct the raw audio. Recently, La-VocE \cite{la_voce} was proposed, to perform AVSE in the mel spectrogram domain in a two-step process: firstly, the ResNet+Transformer-based generator predicts the enhanced mel spectrograms and secondly, a HiFiGAN\cite{hifi_gan} vocoder (trained from scratch) transforms them into the final raw waveforms. This approach demonstrated superior results across multiple noise sources and languages.

\section{Raw waveform models}
\label{section:av2a_raw_waveform_models}

\begin{figure*}
  \includegraphics[width=\linewidth]{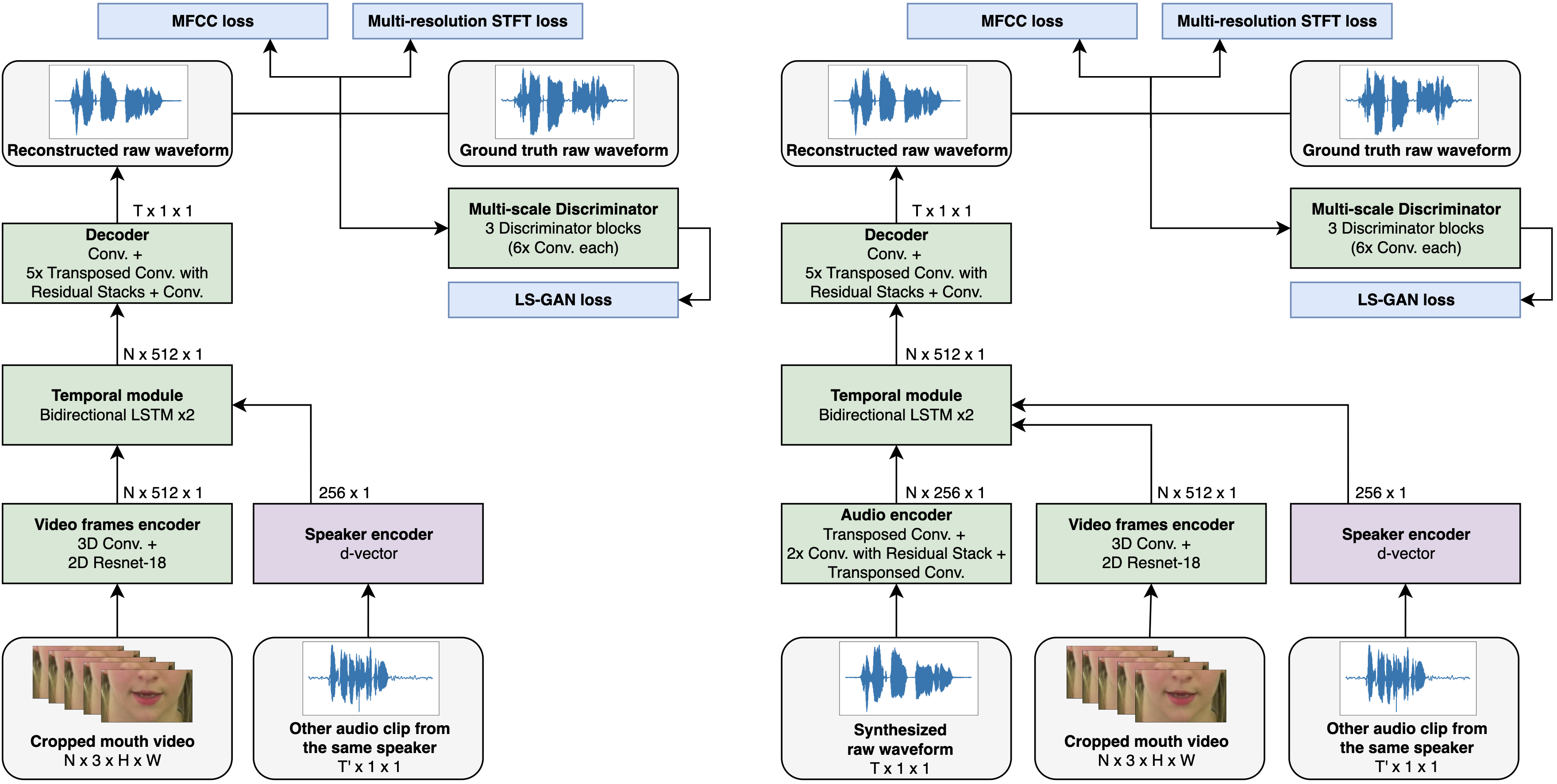}
\caption{(a) V2A-WaveGAN: Video-to-audio model with raw waveforms \cite{large_scale_unsupervised_audio_pretraining_for_v2a}
\hspace{0.2cm} (b) AV2A-WaveGAN: Corresponding Audio-visual-to-audio model \\
Green modules are trainable, purple modules are pre-trained and frozen}
\label{fig:av2a_raw_waveform_high_level}
\end{figure*}

Fig. \ref{fig:av2a_raw_waveform_high_level} shows an overview of the structure of the video-to-audio models we are considering in the raw waveform domain. We employ V2A-WaveGAN as the base model, introduced in our previous work\cite{large_scale_unsupervised_audio_pretraining_for_v2a} and illustrated in Fig. \ref{fig:av2a_raw_waveform_high_level}(a). It is composed of a Generator (a video frames encoder and a speaker encoder, followed by a bidirectional LSTM and a convolutional decoder) and a convolutional Discriminator, to improve the realism of the reconstructed waveforms. The video frames encoder is based on the ResNet-18 architecture (as in \cite{deep_residual_learning_for_image_recognition, visual_speech_recognition_for_multiple_languages_in_the_wild, end_to_end_video_to_speech_synthesis_using_gans}) and the speaker encoder is pre-trained on large-scale speech corpora\cite{real_time_voice_cloning}. We then construct the corresponding audio-visual-to-audio model, AV2A-WaveGAN, by adding an audio encoder to receive the synthesized raw waveforms and by modifying the input dimensionality of the temporal module.

\subsection{Video frames encoder}

\begin{figure}
  \includegraphics[width=\linewidth]{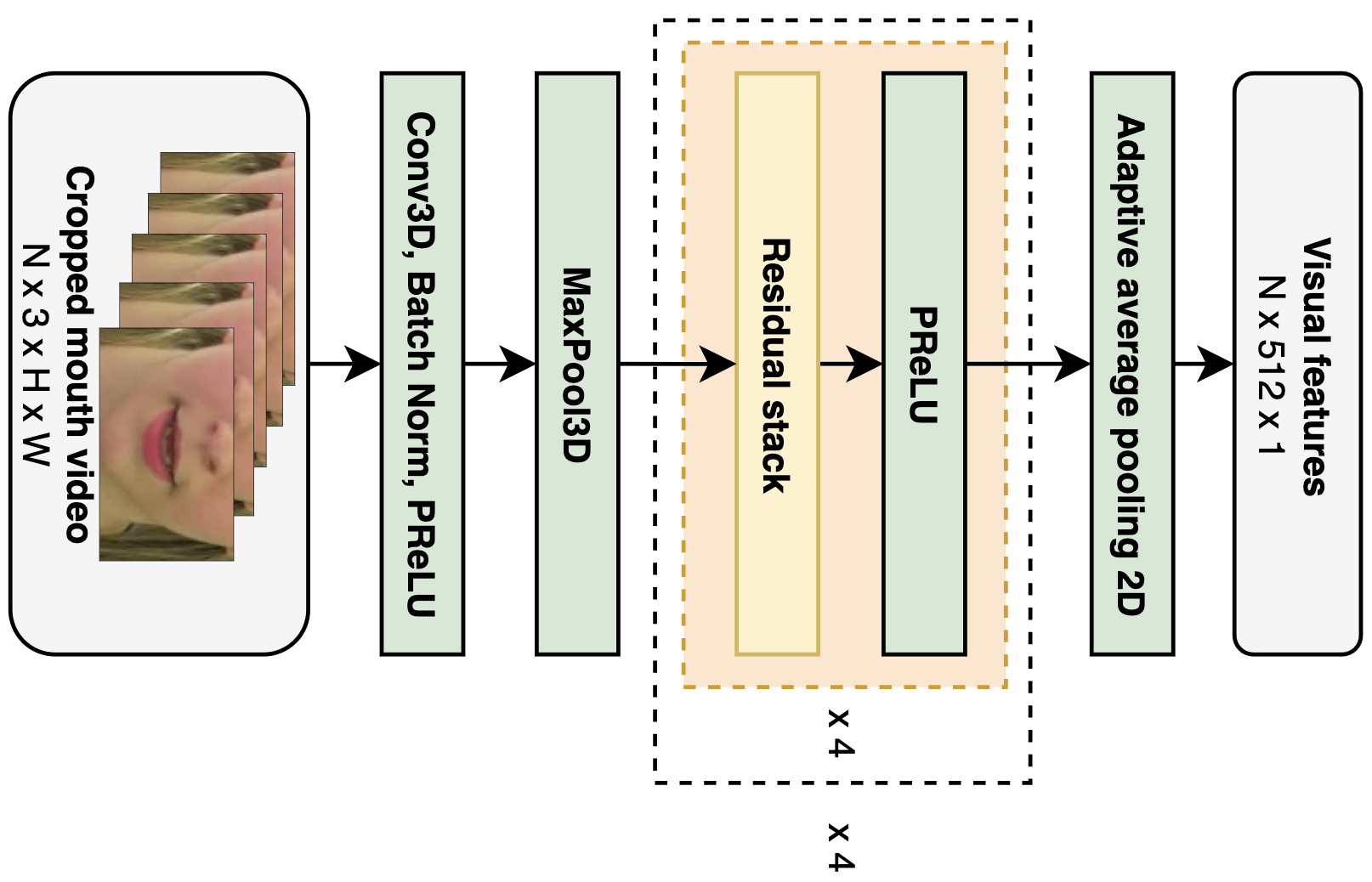} 
\caption{Video frames encoder \cite{large_scale_unsupervised_audio_pretraining_for_v2a}}
\label{video_frames_encoder_simplified}
\end{figure}

To encode the RGB video frames (Fig. \ref{video_frames_encoder_simplified}) we use a 3D spatio-temporal convolution layer, with batch normalization, a PReLU (Parametric Rectified Linear Unit) activation function and max pooling. By using a receptive field of 5 frames, and centering on the current video frame (timestep) this layer has a temporal context of 2 future and 2 past frames. This is followed by a 2D Resnet-18 comprised of 4 blocks of 4 convolutional residual stacks each. Finally, the extracted features are passed to an adaptive average pooling layer. An input video frame of arbitrary height and width is thus encoded into a $512$-D vector.

\subsection{Speaker encoder}
We include a speaker identity module that encodes information about the speaker's biometric characteristics, following \cite{svts, large_scale_unsupervised_audio_pretraining_for_v2a}, and inspired by multi-speaker TTS\cite{transfer_learning_from_speaker_verification}. As in \cite{svts, large_scale_unsupervised_audio_pretraining_for_v2a}, we encode a speaker's voice using the pre-trained speaker encoder of \cite{real_time_voice_cloning, real_time_voice_cloning_github}, trained on the task of speaker verification on VoxCeleb1\cite{voxceleb1_paper}, VoxCeleb2 \cite{voxceleb2_paper} and LibriSpeech \cite{librispeech}. The resulting $256$-D representation is known as a d-vector. To compute a speaker embedding for a given input video, an audio sample from the same speaker (but corresponding to a different video) is selected at random and encoded into a d-vector. The parameters of the speaker encoder are frozen during training.

\subsection{Video-to-audio Generator}
\label{subsection:av2a_raw_waveform_models_v2a_generator}

Given an input video of $N$ frames and a randomly selected audio sample from the same speaker, the video frames and speaker encoders produce a sequence of $N$ $512$-D visual features and a $256$-D speaker embedding respectively. The speaker embedding is then concatenated at each video timestep resulting in a sequence of $N$ $768$-D features passed as input to the temporal module. This consists of a 2-layer bidirectional LSTM and outputs a sequence of $N$ $512$-D temporal features.

\begin{figure}[t]
  \includegraphics[width=\linewidth]{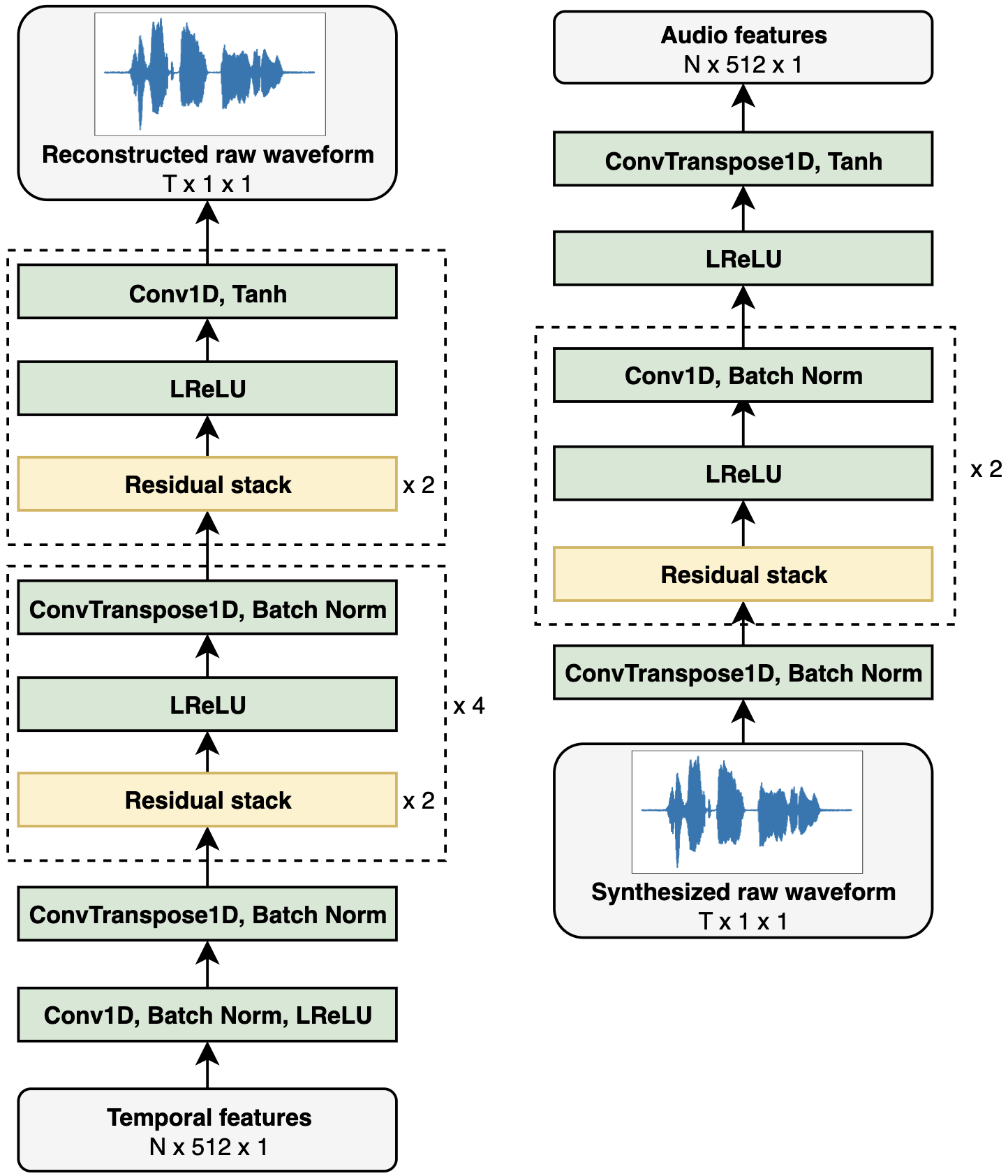}
\caption{Raw waveform decoder (left)\cite{large_scale_unsupervised_audio_pretraining_for_v2a} and audio encoder (right)}
\label{av2a_raw_waveform_encoder_and_decoder_simplified}
\end{figure}

The temporal features are then fed to the decoder, consisting of a convolution, a transposed convolution block and 5 residual blocks followed by a final tanh activation function. Each residual block contains two residual stacks followed by Leaky ReLU and a convolution block (Fig. \ref{av2a_raw_waveform_encoder_and_decoder_simplified}). To imprint an inductive bias of temporal correlation in the audio signal\cite{melgan} we employ dilated convolutions\cite{wavenet} in the residual stacks. The dilation factor increases with increasing number of stacks, in order to increase the induced receptive field for each output timestep. Additionally, the convolution blocks upsample their input, ensuring that each residual block progressively increases the temporal resolution. 

Given a sequence of $N$ temporal features the decoder upsamples them to a raw waveform of $T$ timesteps. Thus, with an audio-video pair sampled at a video frame rate of 25 frames per second and an audio sampling rate of 24 kHz, the decoder outputs an audio signal of 960 timesteps per video timestep.

\subsection{Audio-visual-to-audio Generator}
\label{subsection:av2a_raw_waveform_models_av2a_generator}

We construct the audio-visual-to-audio Generator by appending an audio encoder to the video-to-audio Generator, which receives the synthesized raw waveforms as input and outputs a lower-dimensional representation, as shown in Fig. \ref{fig:av2a_raw_waveform_high_level} (right). We also increase the input dimensionality of the temporal module to accommodate the increased dimensionality of the concatenated input features.

The audio encoder consists of a transposed convolution layer with batch normalization, followed by two residual blocks, a Leaky ReLU activation and a final transposed convolution layer with the tanh activation. This is inspired from the raw waveform encoder used in \cite{large_scale_unsupervised_audio_pretraining_for_v2a} which used 5 residual blocks compared to 2 in this work. Given an input raw waveform sampled at 24 kHz the audio encoder produces a sequence of $256$-D features at a rate of 25 features per second, in line with the sampling rate of the video in our experiments.

Given an input video of $N$ frames, sampled at 25 frames per second, the corresponding synthesized raw waveform of $T$ timesteps, sampled at 24 kHz, the video frames encoder and audio encoder produce sequences of $N$ $512$-D and $256$-D features respectively. These are concatenated, along with an extracted $256$-D speaker embedding at each timestep, to produce an $N$ $1024$-D sequence of features. As with V2A-WaveGAN\cite{large_scale_unsupervised_audio_pretraining_for_v2a}, the temporal module is a 2-layer bidirectional LSTM, modified with input dimensionality of 1024, producing a sequence of $N$ $512$-D features. The decoder remains the same as in the video-to-audio model. 

\subsection{Discriminator}
\label{subsubsection:av2a_raw_waveform_generation_discriminator}

\begin{figure}
  \includegraphics[width=\columnwidth]{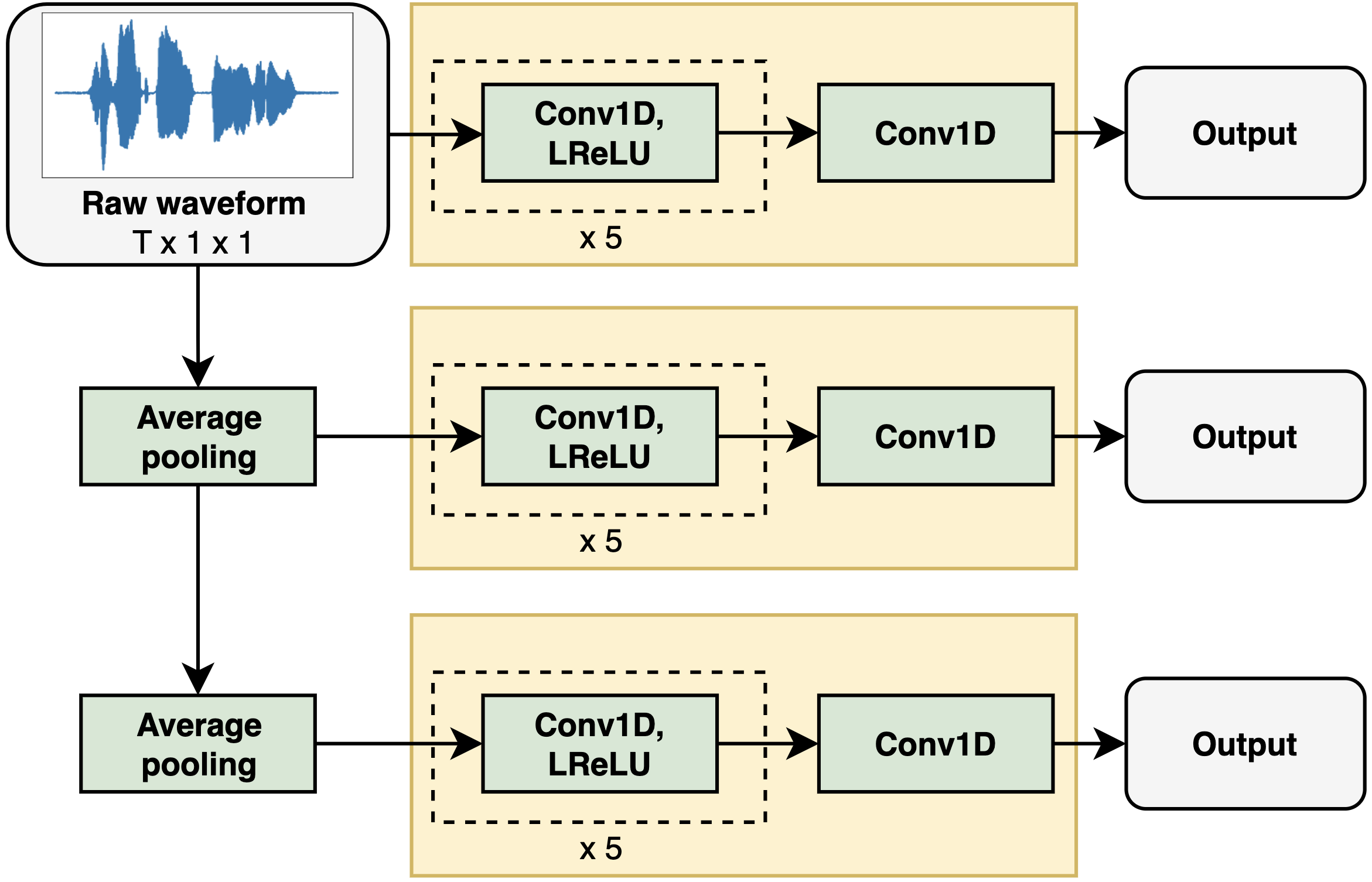}
\caption{Multi-scale discriminator for the raw waveform \cite{melgan, large_scale_unsupervised_audio_pretraining_for_v2a}. Each discriminator block is shaded in yellow.}
\label{av2a_raw_waveform_discriminator}
\end{figure}

We employ the multi-scale discriminator architecture of MelGAN \cite{melgan} (Fig. \ref{av2a_raw_waveform_discriminator}) as in \cite{large_scale_unsupervised_audio_pretraining_for_v2a}, for both V2A-WaveGAN and AV2A-WaveGAN. The Discriminator contains 3 networks of identical architecture, where each network computes a low-dimensional representation of an input waveform at some given scale.

The first discriminator operates at the scale of the original waveform, while we downsample by 2x for each subsequent discriminator. By employing multiple discriminators at different scales one may capture structures in raw audio that are present in different frequencies \cite{melgan}. In addition, it has been observed that using only one discriminator on raw waveforms results in the generator producing metallic audio \cite{melgan}. Following \cite{melgan} we apply weight normalization to all Discriminator layers.

\subsection{Loss function}
\label{subsection:av2a_raw_waveform_generation_loss_function}
Both V2A-WaveGAN and AV2A-WaveGAN are trained using the LS-GAN loss \cite{lsgan}, as in \cite{large_scale_unsupervised_audio_pretraining_for_v2a}, defined as follows for the Generator and the multi-scale discriminator respectively:

\begin{equation}
\label{eq:gen_adv_loss}
L_G = \mathbb{E}_{\mathbf{\Tilde{x}} \sim \mathbb{P}_G}\bigg{[}\sum_{k = 1}^{K}(D_k(\mathbf{\Tilde{x}})-1)^2\bigg{]}
\end{equation}

\begin{equation}
\label{eq:disc_adv_loss}
L_D = \mathbb{E}_{\mathbf{x} \sim \mathbb{P}_X}\bigg{[}\sum_{k = 1}^{K} (D_k(\mathbf{x})-1)^2\bigg{]} + \mathbb{E}_{\mathbf{\Tilde{x}} \sim \mathbb{P}_G}\bigg{[}\sum_{k = 1}^{K}D_k(\mathbf{\Tilde{x}})^2\bigg{]}
\end{equation}

\noindent
where $G$ is the Generator, $D$ is the multi-scale discriminator, $D_k$ is the $k$th discriminator for $k=1, 2, ... K$ scales, $\mathbf{x} \sim \mathbb{P}_X$ are samples from the data distribution and $\mathbf{\Tilde{x}} \sim \mathbb{P}_G$ are samples from the Generator's distribution.

In addition, we include two reconstruction losses to train the Generator \cite{large_scale_unsupervised_audio_pretraining_for_v2a}. The first is the multi-resolution STFT loss \cite{parallel_wavegan}. The STFT loss for a single resolution, $L_S$, is defined as:

\begin{gather}
\label{eq:single_stft_loss}
L_S(\mathbf{x}, \mathbf{\Tilde{x}}) = L_{SC}(\mathbf{x}, \mathbf{\Tilde{x}}) + L_{MAG}(\mathbf{x}, \mathbf{\Tilde{x}}) \\
L_{SC}(\mathbf{x}, \mathbf{\Tilde{x}}) = \frac{|| \ |STFT(\mathbf{x})| \ ||_F - || \ |STFT(\mathbf{\Tilde{x}})| \ ||_F}{|| \ |STFT(\mathbf{x})| \ ||_F} \\
L_{MAG}(\mathbf{x}, \mathbf{\Tilde{x}}) = \frac{1}{n}|| \ log|STFT(\mathbf{x})| - log|STFT(\Tilde{\mathbf{x}})| \ ||_1
\end{gather}

\noindent
and consists of the spectral convergence loss $L_{SC}$ and the log-STFT magnitude loss $L_{MAG}$, where $||\cdot||_F$ $||\cdot||_1$ are the Frobenius and L1 norms respectively and $n$ is the number of elements in the spectrogram.

By combining $M$ STFT losses with different analysis parameters (e.g. FFT size, window size, hop size) one obtains the multi-resolution STFT loss:

\begin{equation}
\label{eq:multi_resolution_stft_loss}
L_{MR\_ STFT}(\mathbf{x}, \mathbf{\Tilde{x}}) = \frac{1}{M}\sum_{m=1}^{M}L_S^{(m)}(\mathbf{x}, \mathbf{\Tilde{x}})
\end{equation}

\noindent
where the $m = 1, 2, ..., M$ denotes the $m$th set of STFT analysis parameters.

The second reconstruction loss is the MFCC loss, introduced in \cite{end_to_end_video_to_speech_synthesis_using_gans}, which aims to increase intelligibility and accuracy of the generated speech as MFCCs (mel-frequency cepstral coefficients \cite{mfcc_paper}) are often used in speech and emotion recognition \cite{end_to_end_video_to_speech_synthesis_using_gans}. It is defined as:

\begin{equation}
\label{eq:mfcc_loss}
L_{MFCC}(\mathbf{x}, \mathbf{\Tilde{x}}) = ||MFCC(\mathbf{x}) - MFCC(\mathbf{\Tilde{x}})||_1
\end{equation}

\noindent
where the $MFCC$ function extracts 25 MFCCs from the raw waveform.

Finally, the Generator's loss function combines the adversarial loss with the aforementioned two reconstruction losses:

\begin{equation}
\label{eq:v2a_wavegan_generator_loss}
L_{GEN} = \lambda_1 L_G + \lambda_2 L_{MR\_ STFT} + \lambda_3 L_{MFCC}
\end{equation}

\noindent
where $\lambda_1, \lambda_2, \lambda_3 >0$ are hyperparameters.

In \cite{large_scale_unsupervised_audio_pretraining_for_v2a}, and following \cite{end_to_end_video_to_speech_synthesis_using_gans}, we tune these coefficients sequentially by incrementally finding the values that yield the lowest word error rate on the validation set of GRID (4 speakers, seen). This yields $\lambda_1 = 1.0$, $\lambda_2 = 80.0$, $\lambda_3 = 15.0$. As these coefficients were employed in V2A-WaveGAN, we thus employ them in our AV2A-WaveGAN experiments as well.

\section{Mel spectrogram models}
\begin{figure*}
  \includegraphics[width=\linewidth]{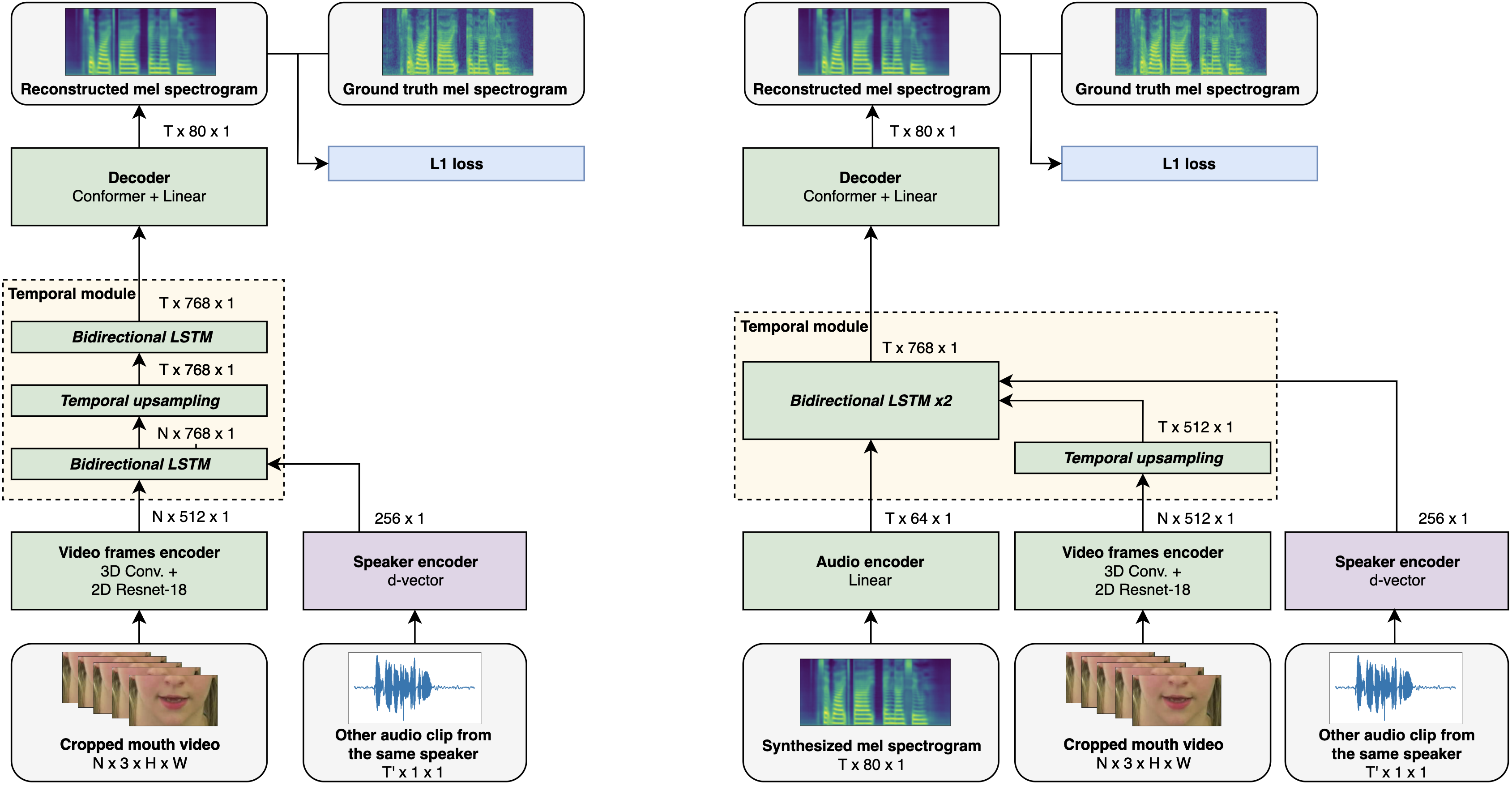}
\caption{(a) V2A-MelSpec: Video-to-audio model with mel spectrograms \cite{large_scale_unsupervised_audio_pretraining_for_v2a}
\hspace{0.2cm} (b) AV2A-MelSpec: Corresponding Audio-visual-to-audio model \\
Green modules are trainable, purple modules are pre-trained and frozen}
\label{fig:av2a_melspec_high_level}
\end{figure*}

Fig. \ref{fig:av2a_melspec_high_level} shows an overview of the structure of the video-to-audio models we are considering in the mel spectrogram domain. We employ V2A-MelSpec as the base model, introduced in our previous work\cite{large_scale_unsupervised_audio_pretraining_for_v2a}. It contains a Generator composed of a video frames encoder and a speaker encoder, followed by a temporal module that upsamples across the temporal axis and a decoder that outputs mel spectrograms. The video frames encoder and speaker encoder consist of the same architectures as in the raw waveform generators in Section \ref{section:av2a_raw_waveform_models}. The temporal module is composed of an LSTM - temporal upsampling - LSTM sequence of layers and the decoder consists of a conformer-based model and a linear projection. We  construct the corresponding audio-visual-to-audio model, AV2A-MelSpec, by adding an audio encoder (one Linear layer) that is fed the synthesized mel spectrograms. In addition, as in Section \ref{section:av2a_raw_waveform_models}, we modify the temporal module to account for the new dimensionality of the inputs as well as the different temporal resolutions of the mel spectrogram and the video frames.

\subsection{Video-to-audio Generator}

As with raw waveform generation in Section \ref{section:av2a_raw_waveform_models}, the video frames encoder receives an input video of $N$ frames and outputs a sequence of $N$ $512$-D visual features. The speaker encoder also produces a $256$-D speaker embedding from another audio sample of the same speaker selected at random. By concatenating the speaker embedding with the visual features at each timestep the resulting sequence of $N$ $768$-D features is fed to the temporal module. The latter is composed of a 1-layer bidirectional LSTM followed by temporal upsampling and another bidirectional LSTM and outputs a sequence of $N$ $768$-D temporal features.

\begin{figure}
  \includegraphics[width=\linewidth]{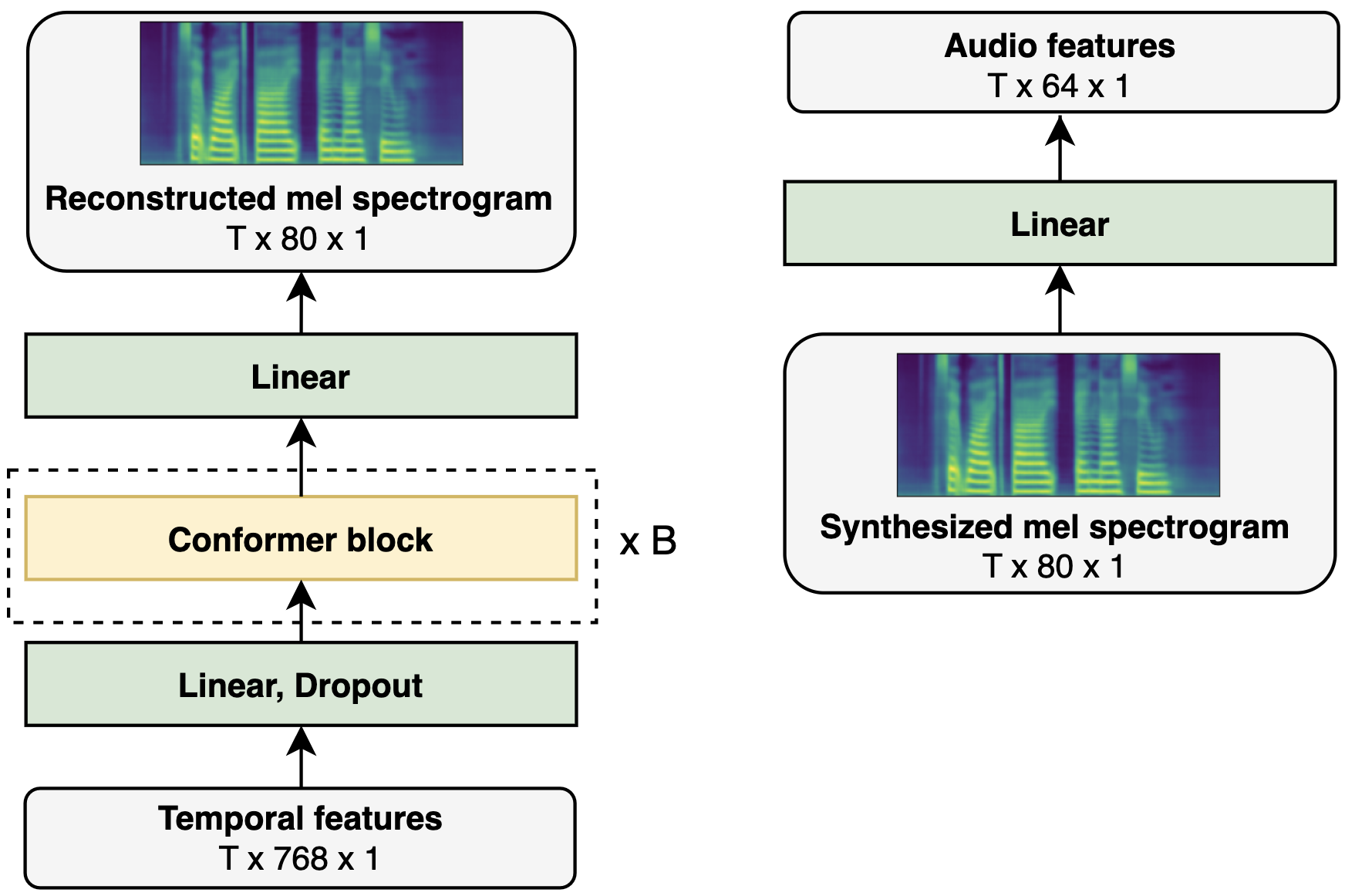}
\caption{Mel spectrogram decoder (left)\cite{large_scale_unsupervised_audio_pretraining_for_v2a} and audio encoder (right).}
\label{av2a_melspec_encoder_and_decoder_simplified}
\end{figure}

The decoder, consisting of a Linear layer followed by dropout, $B$ conformer blocks and a final linear layer, receives the temporal features and outputs a mel spectrogram. Fig. \ref{av2a_melspec_encoder_and_decoder_simplified} (left) and Table \ref{table:av2a_melspec_decoder_architectures} illustrate the decoder architecture and the conformer block configurations. Inspired by the originally proposed Conformer \cite{conformer} and in line with with previous works \cite{svts, large_scale_unsupervised_audio_pretraining_for_v2a}, the number of conformer blocks $B$ is set as a hyperparameter in proportion to the size of the training dataset. This results in three versions of the decoder, and thus the Generator.

\begin{table}[h]
\captionsetup{justification=centering}
\caption{Summary of V2A-MelSpec / AV2A-MelSpec \\ decoder architectures}
\begin{adjustbox}{width=\columnwidth}
\begin{tabular}{@{}lccc@{}}
\toprule
\multicolumn{1}{c}{Model} & \begin{tabular}[c]{@{}c@{}}V2A-MelSpec \slash\\AV2A-MelSpec\\ (VS)\end{tabular} & \begin{tabular}[c]{@{}c@{}}V2A-MelSpec \slash\\AV2A-MelSpec\\ (S)\end{tabular} & \begin{tabular}[c]{@{}c@{}}V2A-MelSpec \slash\\AV2A-MelSpec\\ (M)\end{tabular} \\ \midrule
Conformer blocks          & 2                                                          & 6                                                         & 12                                                        \\
Attention dim.            & 256                                                        & 256                                                       & 256                                                       \\
Attention heads           & 4                                                          & 4                                                         & 4                                                         \\
Conv. kernel size         & 31                                                         & 31                                                        & 31                                                        \\
Feedforward dim.          & 2048                                                       & 2048                                                      & 2048                                                      \\ \bottomrule
\end{tabular}
\end{adjustbox}
\label{table:av2a_melspec_decoder_architectures}
\end{table}

\subsection{Audio-visual-to-audio Generator}
To construct the audio-visual-to-audio Generator we append a Linear layer as the audio encoder (following \cite{av_hubert, la_voce}), to the video-to-audio Generator. This receives synthesized mel spectrograms as input and produces a lower-dimensional representation. By using a simple, lightweight audio encoder we prevent the Generator from relying excessively on the audio input \cite{av_hubert}, i.e. from trivially ignoring the visual modality and learning an identity mapping.

Given a synthesized mel spectrogram of $T$ timesteps and $80$ frequency bins, the audio encoder outputs a sequence of $T$ $64$-D features. As the video frames encoder outputs a sequence of $N$ $512$-D features (corresponding to $N$ video frames, sampled at 25 frames per second, $N<T$), we need to align the temporal resolutions of the audio and video representations prior to concatenating them. Thus, we upsample the visual features from $N$ to $T$ timesteps using nearest-neighbor upsampling along the time axis. We then concatenate these as well as the $256$-D extracted speaker embedding at each timestep, resulting in a sequence of $T$ $832$-D features. These are then fed to a bidirectional LSTM outputting a sequence of $T$ $768$-D temporal features to be fed to the decoder. The decoder has the same architecture as in the video-to-audio model.

To generate the raw waveforms from the reconstructed mel spectrograms we use HiFiGAN\cite{hifi_gan}, a neural vocoder pre-trained on the LibriTTS\cite{libri_tts} dataset containing more than 500 hours of speech data from audiobooks. The pre-trained model is publicly available\footnotemark.

\footnotetext{\texttt{\url{https://github.com/kan-bayashi/ParallelWaveGAN}}}

\subsection{Loss function}
\label{subsection:av2a_melspec_generation_loss_function}

We employ an L1 loss on the mel spectrograms \cite{large_scale_unsupervised_audio_pretraining_for_v2a}:

\begin{equation}
\label{eq:v2a_melspec_generator_loss}
L_{GEN}(\mathbf{X}, \mathbf{\Tilde{X}}) = || \mathbf{X} - \mathbf{\Tilde{X}} ||_1
\end{equation}

\noindent
where $\mathbf{X}$ and $\mathbf{\Tilde{X}}$ denote the ground truth and generated mel spectrograms, respectively.

\section{Experimental methodology and setup}
\subsection{Datasets}
\label{subsection:av2a_datasets}

We conduct experiments on three audio-visual face and speech datasets which are widely used in the video-to-speech literature: GRID \cite{grid_database}, TCD-TIMIT \cite{tcd_timit_databse} and LRW\cite{lip_reading_in_the_wild}. These are summarized in Table \ref{table:av2a_av_datasets}.

GRID \cite{grid_database} is composed of 33 speakers that utter structured sentences. Each sentence is recorded in laboratory conditions and contains 6 words selected at random from a fixed vocabulary of 51 words. In line with the literature, we experiment with three versions of this dataset: (1) a seen speaker setting with 4 speakers, used in several previous works \cite{end_to_end_video_to_speech_synthesis_using_gans, video_driven_speech_reconstruction_using_gans, lip2wav, lip_to_speech_synthesis_with_visual_context_attention_gan, large_scale_unsupervised_audio_pretraining_for_v2a}; (2) a seen speaker setting with 33 speakers, originally proposed in \cite{end_to_end_video_to_speech_synthesis_using_gans} and (3) an unseen speaker setting with 33 speakers used in \cite{end_to_end_video_to_speech_synthesis_using_gans, video_driven_speech_reconstruction_using_gans, lip_to_speech_synthesis_with_visual_context_attention_gan, large_scale_unsupervised_audio_pretraining_for_v2a}.

TCD-TIMIT \cite{tcd_timit_databse} consists of 62 speakers (59 volunteers and 3 lipspeakers) uttering sentences recorded in laboratory conditions. In line with previous works \cite{lip2wav, end_to_end_video_to_speech_synthesis_using_gans, lip_to_speech_synthesis_with_visual_context_attention_gan, large_scale_unsupervised_audio_pretraining_for_v2a} we use the audio and video samples of the 3 lipspeakers only, which amount to 377 clips per lipspeaker. We employ a seen speaker split proposed in \cite{end_to_end_video_to_speech_synthesis_using_gans}.

LRW \cite{lip_reading_in_the_wild} consists of 'in-the-wild' audio-visual clips of one word utterances from hundreds of speakers, out of a vocabulary of 500 words. These samples were obtained from television shows and include variations in lighting conditions, head pose and background noise. This dataset is therefore more challenging for face and speech-related tasks than GRID and TCD-TIMIT.

\begin{table}[h]
\captionsetup{justification=centering}
\caption{Summary of audio-visual datasets}
\begin{adjustbox}{width=\columnwidth}
\begin{tabular}{@{}lccc@{}}
\toprule
\textbf{Dataset}                & \textbf{\begin{tabular}[c]{@{}c@{}}Training set\\ (samples/hours)\end{tabular}} & \textbf{\begin{tabular}[c]{@{}c@{}}Validation set\\ (samples/hours)\end{tabular}} & \textbf{\begin{tabular}[c]{@{}c@{}}Test set\\ (samples/hours)\end{tabular}} \\ \midrule
GRID (4 speakers, seen)         & 3543 / 2.95                                                                              & 209 / 0.17                                                                                 & 209 / 0.17                                                                           \\
GRID (33 speakers, seen)        & 29333 / 24.44                                                                              & 1627 / 1.36                                                                                 & 1630 / 1.36                                                                          \\
GRID (33 speakers, unseen)      & 15648 / 13.04                                                                              & 6996 / 5.83                                                                                 & 9946 / 8.29                                                                          \\
TCD-TIMIT (3 lipspeakers, seen) & 1014 / 1.64                                                                              & 57 / 0.09                                                                                 & 60 / 0.10                                                                           \\
LRW (unseen)                    & 480456 / 154.81                                                                              & 24728 / 7.97                                                                                & 24663 / 7.95                                                                          \\ \bottomrule
\end{tabular}
\end{adjustbox}
\label{table:av2a_av_datasets}
\end{table}

\subsection{Data pre-processing}
We extract cropped mouth regions from the video frames by first conducting face detection using the S$^3$FD \cite{s3fd}, followed by landmark localization of 68 points with a pre-trained 2D-FAN \cite{fan} and alignment of each face to a reference mean face. We then crop mouth regions of dimensionality $128\times74$ for GRID and $96\times96$ for TCD-TIMIT and LRW in line with \cite{end_to_end_video_to_speech_synthesis_using_gans} and normalize the resulting images. We also employ data augmentation during training by applying horizontal flipping to each cropped frame with a 50\% probability.

All audio is sampled at 24 kHz. To extract log-mel spectrograms we use 80 mel bands, an FFT size of 2048, a hop size of 12.5 ms, a window length of 50 ms, and a Hann window. Furthermore we clip values outside the $[-6, 6]$ range and rescale the resulting spectrogram to $[-1, 1]$. We do not pre-process the raw waveforms.

\subsection{AV2A model training}
Given a pre-trained V2A model, we begin by synthesizing the audio (raw waveforms or mel spectrograms) of a dataset of interest, including the training, validation and test subsets. Then, in the corresponding AV2A model, we initialize the video frames encoder and the decoder with the parameters of the pre-trained V2A model, and use default initialization for the audio encoder and the temporal module. If the V2A model includes a Discriminator, as in V2A-WaveGAN, we initialize the Discriminator of the the corresponding AV2A model with the pre-trained parameters.

For a given pre-trained V2A model, we experiment with three training procedures with the corresponding AV2A model:

\begin{itemize}
\item \textbf{Baseline:} The temporal module receives a sequence of features from the audio and video frames encoders (including the speaker embedding), i.e. from both audio and visual modalities. 
\item \textbf{Modality dropout:} This consists of alternating between reconstructing speech from all modalities (as in the baseline), from the visual modality only (including the speaker embedding), and from the audio modality only. This is described in Algorithm \ref{algorithm:av2a_training_with_modality_dropout} where $E_A, E_V, E_I$ denote the audio encoder, video frames encoder and speaker encoder respectively, $T_{AV}$ and $F_{AV}$ denote the temporal module and decoder and $D_{AV}$ is the Discriminator (included only if part of the model).
\item \textbf{Modality dropout (ground truth audio): } This involves performing modality dropout, as above, whereby during reconstruction with the audio modality only, we compute audio features from the ground truth audio rather than the synthesized audio. When reconstructing speech from all modalities we keep using the synthesized audio as input to the audio decoder, as with modality dropout above. 
\end{itemize}

\subsubsection{Selection of pre-trained base V2A models}
For each audio domain (raw waveform and mel spectrograms) and each dataset, we obtain two pre-trained base V2A models from our previous work \cite{large_scale_unsupervised_audio_pretraining_for_v2a}: the first model was trained from scratch (i.e. with random parameter initialization) and the second model was trained with all parameters initialized at random, except of the decoder and optionally the Discriminator. The parameters of the latter were pre-trained on a large volume of audio data, as part of the task of audio auto-encoding which was the subject of our investigation \cite{large_scale_unsupervised_audio_pretraining_for_v2a}. As we proposed multiple methods for fine-tuning V2A models with these pre-trained modules, in this work we select for each audio domain and dataset the fine-tuned model with the lowest validation loss. Thus, for each audio domain and dataset we obtain two pre-trained base V2A models which we note in section \ref{section:av2a_results} in the format V2A-X and V2A-X with audio pre-training.

\begin{algorithm}[t]
\caption{AV2A-WaveGAN and AV2A-MelSpec training with modality dropout}
\begin{algorithmic}[1]
\STATE \textbf{Input:} \texttt{train\_data}, \texttt{ground\_truth\_audio}, \\ $E_A, E_V, E_I, T_{AV}, F_{AV}$, [optional] $D_{AV}$
\STATE \texttt{for epoch in num\_epochs:}
\STATE \quad \texttt{for batch in train\_data:}
\STATE \qquad \textbackslash\textbackslash \: Synthesized audio, video frames, ground truth \\ \qquad audio and input to speaker encoder
\STATE \qquad $\mathbf{x_a}, \mathbf{x_v}, \mathbf{x}, \mathbf{x_I} \gets $ \texttt{batch}
\STATE 
\STATE \qquad \textbackslash\textbackslash \: Reconstruct speech with all modalities
\STATE \qquad $\mathbf{z_a}, \mathbf{z_v}, \mathbf{z_I} = E_A(\mathbf{x_a}), E_V(\mathbf{x_v}), E_I(\mathbf{x_I})$
\STATE \qquad $\mathbf{z} = T_{AV}(\mathbf{z_a}, \mathbf{z_v}, \mathbf{z_I})$
\STATE \qquad $\mathbf{\Tilde{x}} = F_{AV}(\mathbf{z})$
\STATE \qquad \texttt{compute\_loss} $(\mathbf{x}, \mathbf{\Tilde{x}})$
\STATE \qquad \texttt{backpropagation} ($E_A, E_V, T_{AV}, F_{AV}, D_{AV}$)
\STATE
\STATE \qquad \textbackslash\textbackslash \: Reconstruct speech with visual modality only \\ \qquad (including speaker embedding)
\STATE \qquad $\mathbf{z_v}, \mathbf{z_I} = E_V(\mathbf{x_v}), E_I(\mathbf{x_I})$
\STATE \qquad $\mathbf{z} = T_{AV}(\mathbf{0}, \mathbf{z_v}, \mathbf{z_I})$
\STATE \qquad $\mathbf{\Tilde{x}} = F_{AV}(\mathbf{z})$
\STATE \qquad \texttt{compute\_loss} $(\mathbf{x}, \mathbf{\Tilde{x}})$
\STATE \qquad \texttt{backpropagation} ($E_V, T_{AV}, F_{AV}, D_{AV}$)
\STATE 
\STATE \qquad \textbackslash\textbackslash \: Reconstruct speech with audio modality only
\STATE \qquad \textbf{if} \texttt{ground\_truth\_audio} \textbf{ is True:}
\STATE \qquad \qquad $\mathbf{z_a} = E_A(\mathbf{x})$
\STATE \qquad \textbf{else:}
\STATE \qquad \qquad $\mathbf{z_a} = E_A(\mathbf{x_a})$
\STATE \qquad $\mathbf{z} = T_{AV}(\mathbf{z_a}, \mathbf{0}, \mathbf{0})$
\STATE \qquad $\mathbf{\Tilde{x}} = F_{AV}(\mathbf{z})$
\STATE \qquad \texttt{compute\_loss} $(\mathbf{x}, \mathbf{\Tilde{x}})$
\STATE \qquad \texttt{backpropagation} ($E_A, T_{AV}, F_{AV}, D_{AV}$)
\STATE \textbf{Output:} $E_A, E_V, E_I, T_{AV}, F_{AV}$, [optional] $D_{AV}$
\end{algorithmic}
\label{algorithm:av2a_training_with_modality_dropout}
\end{algorithm}

\subsubsection{Raw waveform training details}
We train the Generator and Discriminator of AV2A-WaveGAN using a batch size of 4 and with the Adam \cite{adam} optimizer with a learning rate of $1 \times 10^{-4}$, $\beta_1 = 0.5$ and $\beta_2 = 0.99$. In line with \cite{end_to_end_video_to_speech_synthesis_using_gans} the Discriminator receives a 1 second audio clip cropped at random from the ground truth and reconstructed audio. The Generator is fed input samples of duration up to 3 seconds.

\subsubsection{Mel spectrogram training details}
We train AV2A-MelSpec using the AdamW \cite{adamw} optimizer with $\beta_1 = 0.9, \beta_2 = 0.98$ and weight decay of $1 \times 10^{-2}$. Training is conducted in two stages:
\begin{itemize}
\item \textbf{Training audio encoder and temporal module only:} The video frames encoder and the decoder are kept frozen. The remaining trainable modules (audio encoder and Bidirectional LSTM in the temporal module) are trained using a learning rate of $1 \times 10^{-3}$ for all experiments with seen speakers and $5 \times 10^{-4}$ for all experiments with unseen speakers. We warmup the learning rate for 20 epochs, for experiments with GRID and TCD-TIMIT, and for 15 epochs for experiments with LRW. Following the warmup phase, we decay the learning rate using a cosine schedule with warm restarts\cite{sgdr}, where $T_0 = 1, T_{mult} = 2$. We save a checkpoint at the end of each epoch.
\item \textbf{Continuing training the entire model:} From the above we select the checkpoint of the epoch with the lowest validation loss during the warmup phase and un-freeze the video frames encoder and decoder. We continue training the model with the given learning rate schedule, where the learning rates for the audio encoder and temporal module are as above and those of the video frames encoder and decoder are set to $1 \times 10^{-4}$ for experiments with seen speakers and $1 \times 10^{-5}$ for experiments with unseen speakers.
\end{itemize}

\subsection{Evaluation metrics}
To measure the quality and intelligibility of our reconstructed speech we employ 4 objective metrics widely used in speech synthesis works: PESQ, STOI, ESTOI  and the word error rate (WER). Although it is widely acknowledged that existing metrics correlate imperfectly with human perception \cite{video_driven_speech_reconstruction_using_gans, end_to_end_video_to_speech_synthesis_using_gans}, these metrics are widely used in the video-to-speech literature and are useful for comparing different works.

PESQ (perceptual evaluation of speech quality)\cite{pesq}, originally created to measure speech quality of speech codecs and telephone networks, aims to capture the perceptual quality of the speech. STOI (short-time objective intelligibility)\cite{stoi} and its extended version ESTOI \cite{estoi} aim to measure the intelligibility of speech samples and show high correlation with reported intelligibility scores in subjective listening tests. For these three metrics, higher scores are better.

WER measures the word-level accuracy of speech samples. Following common practice, we use pre-trained speech recognition models to compute it. For experiments with GRID we use a pre-trained model \cite{visual_speech_recognition_for_multiple_languages_in_the_wild, svts} which achieves a WER of 0.1\% on the real audio test set (following the split in \cite{lipnet}). For LRW we use a model trained on LRW \cite{end_to_end_av_speech_recognition} and achieving a WER of 1.68\% on the test set. We did not calculate WER for TCD-TIMIT as we were unable to find an accurate, publicly available speech recognition model for it.

\section{Results}
\label{section:av2a_results}

This section presents our experiments using the datasets and splits defined in section \ref{subsection:av2a_datasets} and Table \ref{table:av2a_av_datasets}, with the results shown in Tables \ref{table:av2a_grid_4_seen_speakers} - \ref{table:av2a_lrw}. We benchmark our models with comparable methods in the same audio domain, i.e. AV2A-WaveGAN models with other raw waveform methods and AV2A-MelSpec models with other methods generating acoustic features (such as mel spectrograms). Within each domain, we group the results of the base V2A model with those of our corresponding AV2A experiments to make explicit the impact of AV2A over the base V2A model. For all methods, we report results on test set samples provided by their respective authors, with the exception of Lip2Wav\cite{lip2wav} where results are shown as stated in their paper. Our generated samples can be viewed on our project website \footnotemark.

\footnotetext{\texttt{\url{https://sites.google.com/view/v2a-audio-visual/home}}}

We observe that our raw waveform models outperform all other comparable methods across all metrics, except for WER in GRID (33 speakers, unseen). We observe that employing modality dropout with ground truth audio outperforms both the base V2A-WaveGAN and AV2A-WaveGAN. Furthermore, this holds both when the base V2A model is trained from scratch and when audio pre-training is used. Across all datasets, except with GRID (33 speakers, unseen), modality dropout with ground truth audio produces better results than modality dropout. In most cases we also see that when the base V2A model is trained with audio pre-training, the corresponding AV2A models outperform the AV2A models where the base V2A model was trained from scratch. This suggests that in those cases the advantage of audio pre-training in V2A is maintained when fine-tuning the shared modules (video frames encoder, decoder and Discriminator) in AV2A.

Within acoustic features models, we observe that our mel spectrogram models produce improved metrics compared to other works in the majority of cases. Across all datasets, except TCD-TIMIT, our AV2A models show higher reconstruction metrics than their base V2A models, as well as lower WER in most cases. However, unlike our AV2A models in the raw waveform domain, there is no clear winner among our three proposed training procedures, despite there being at least one such method outperforming the base V2A model in every dataset (except TCD-TIMIT). We note that across all datasets, the validation and test set losses in all AV2A-MelSpec models were lower than the corresponding losses of the base V2A-MelSpec model when the latter was trained from scratch. When the base V2A-MelSpec model was trained with audio pre-training, the validation and test set losses were sometimes higher in the corresponding AV2A-MelSpec experiments compared to the base V2A-MelSpec models. For example, this occurs with GRID (33 speakers, unseen), despite the fact that the base V2A-MelSpec model with audio pre-training has both lower loss and better objective metrics than V2A-MelSpec trained from scratch. This suggests that the advantage of audio pre-training in the base model, in the form of lower loss, does not necessarily carry on in the corresponding AV2A model. We conjecture that this is an optimization difficulty and a consequence of the fact that the decoder has been fine-tuned twice: once when training the V2A model (i.e., fine-tuning from the audio pre-training), and a second time in this case when training the corresponding AV2A model. We defer further investigation of this problem to future work.

\subsection{Results on seen speakers}

Our results on experiments using dataset splits with seen speakers are shown in Tables \ref{table:av2a_grid_4_seen_speakers} - \ref{table:av2a_tcd_timit_lipspeakers}. In our experiments with GRID (4 speakers, seen), AV2A-WaveGAN with audio pre-training and modality dropout (with ground truth audio) outperforms all other raw waveform models across all metrics. All three training procedures of AV2A-WaveGAN result to improved reconstruction metrics compared to their base V2A-WaveGAN model, and the WER is equal or lower in most cases. AV2A-MelSpec-VS with audio pre-training improves upon its corresponding base V2A model in most metrics, and achieves the lowest WER among all comparable methods when trained with modality dropout (GT audio). However, VCA-GAN\cite{lip_to_speech_synthesis_with_visual_context_attention_gan} outperforms it in PESQ and Lip2Wav\cite{lip2wav} in STOI and ESTOI.

\begin{table}[h]
\captionsetup{justification=centering}
\caption{Results on GRID (4 speakers, seen)}
\begin{adjustbox}{width=\columnwidth}
\begin{tabular}{@{}lcccc@{}}
\toprule
\textbf{Method}                    & \textbf{PESQ$\uparrow$} & \textbf{STOI$\uparrow$} & \textbf{ESTOI$\uparrow$} & \textbf{WER (\%)$\downarrow$} \\ \midrule \midrule
\textbf{Raw waveform models} \\ \midrule \midrule
End-to-end WGAN (2018) \cite{video_driven_speech_reconstruction_using_gans}                           & 1.47             & 0.570             & 0.329              & 19.94                 \\
End-to-end WGAN (2022) \cite{end_to_end_video_to_speech_synthesis_using_gans}                            & 1.76             & 0.662             & 0.468              & 4.07                 \\

\midrule
V2A-WaveGAN \cite{large_scale_unsupervised_audio_pretraining_for_v2a} & 1.87          & 0.693         & 0.513        & 4.68              \\
\midrule

AV2A-WaveGAN & 1.90          & 0.696         & 0.520        & 4.84              \\
\hspace{3mm} + modality dropout & 1.91          & 0.696         & 0.520        & 4.84              \\
\hspace{3mm} + modality dropout (GT audio)& 1.91          & \textbf{0.698}         & 0.527        & 4.76              \\

\midrule
V2A-WaveGAN with audio pre-training \cite{large_scale_unsupervised_audio_pretraining_for_v2a} & 1.90          & 0.690         & 0.513          & 4.99                       \\
\midrule

AV2A-WaveGAN with audio pre-training & 1.91          & 0.696         & 0.518        & 4.99              \\
\hspace{3mm} + modality dropout & 1.91          & 0.694         & 0.527        & 4.28              \\
\hspace{3mm} + modality dropout (GT audio)& \textbf{1.95}          & \textbf{0.698}         & \textbf{0.532}        & \textbf{3.67}              \\
\midrule \midrule

\textbf{Acoustic features models} \\ \midrule \midrule
Vid2Voc \cite{vocoder_based_speech_synthesis}                            & 1.61             & 0.650             & 0.455              & 9.29                 \\
Lip2Wav \cite{lip2wav}          & 1.77             & \textbf{0.731}             & \textbf{0.535}              & 14.08\footnotemark[2] \\
VCA-GAN \cite{lip_to_speech_synthesis_with_visual_context_attention_gan}                          & \textbf{2.03}             & 0.682             & 0.510  & \textbf{5.62} \\
Visual Voice Memory \cite{speech_reconstruction_visual_voice_memory}          & 1.82             & 0.643             & 0.481              & 6.08 \\

\midrule
V2A-MelSpec-VS \cite{large_scale_unsupervised_audio_pretraining_for_v2a} & 1.83          & 0.693         & 0.505        & 6.70              \\
\midrule
AV2A-MelSpec-VS          & 1.83         & 0.690        & 0.499 & 5.80              \\
\hspace{3mm} + modality dropout & 1.86          & 0.691         & 0.509        & 6.16              \\
\hspace{3mm} + modality dropout (GT audio)& 1.84          & 0.689         & 0.504        & 5.34              \\

\midrule
V2A-MelSpec-VS with audio pre-training \cite{large_scale_unsupervised_audio_pretraining_for_v2a} & 1.87          & 0.695         & 0.512          & 5.74              \\
\midrule
AV2A-MelSpec-VS with audio pre-training & 1.87          & 0.696         & 0.516        & 4.71              \\
\hspace{3mm} + modality dropout & 1.88          & 0.699         & 0.522        & 4.55              \\
\hspace{3mm} + modality dropout (GT audio)& 1.88          & 0.694         & 0.518        & \textbf{4.22}              \\
\bottomrule
\end{tabular}
\end{adjustbox}
\vfill
\hspace{3mm} \\
$^2$Reported using Google speech-to-text API
\label{table:av2a_grid_4_seen_speakers}
\end{table}

\begin{table}[h]
\captionsetup{justification=centering}
\caption{Results on GRID (33 speakers, seen)}
\begin{adjustbox}{width=\columnwidth}
\begin{tabular}{@{}lcccc@{}}
\toprule
\textbf{Method}                    & \textbf{PESQ$\uparrow$} & \textbf{STOI$\uparrow$} & \textbf{ESTOI$\uparrow$} & \textbf{WER (\%)$\downarrow$} \\ \midrule \midrule
\textbf{Raw waveform models} \\ \midrule \midrule
End-to-end WGAN (2022) \cite{end_to_end_video_to_speech_synthesis_using_gans}                            & 1.70             & 0.667             & 0.465              & 4.59                 \\
\midrule
V2A-WaveGAN \cite{large_scale_unsupervised_audio_pretraining_for_v2a} & 2.00          & 0.712         & 0.529        & 2.79              \\
\midrule

AV2A-WaveGAN & 2.05          & 0.716         & 0.537        & 2.81              \\
\hspace{3mm} + modality dropout & 2.03          & 0.715         & 0.536        & 2.78              \\
\hspace{3mm} + modality dropout (GT audio)& 2.06          & 0.717         & 0.544        & \textbf{2.50}              \\

\midrule
V2A-WaveGAN with audio pre-training \cite{large_scale_unsupervised_audio_pretraining_for_v2a}    & 2.07          & 0.716       & 0.539          & 2.83              \\
\midrule

AV2A-WaveGAN with audio pre-training & 2.09          & 0.720         & 0.547        & 2.87              \\
\hspace{3mm} + modality dropout & \textbf{2.10}          & 0.720         & 0.548        & 2.92              \\
\hspace{3mm} + modality dropout (GT audio)& \textbf{2.10}          & \textbf{0.723}         & \textbf{0.553}        & 2.65              \\ \midrule \midrule

\textbf{Acoustic features models} \\ \midrule \midrule
VCA-GAN \cite{lip_to_speech_synthesis_with_visual_context_attention_gan}                            & 1.97             & 0.695             & 0.505              & 5.10                 \\
SVTS-S \cite{svts}                             & 1.97             & 0.705             & 0.523              & \textbf{2.37}                 \\
\midrule
V2A-MelSpec-S \cite{large_scale_unsupervised_audio_pretraining_for_v2a} & 2.02          & 0.720         & 0.538          & 2.66              \\
\midrule
AV2A-MelSpec-S & 2.00          & 0.717         & 0.535        & 2.72              \\
\hspace{3mm} + modality dropout & 2.04          & 0.715         & 0.536        & 2.82              \\
\hspace{3mm} + modality dropout (GT audio)& 2.02          & 0.720         & 0.539        & 2.53              \\

\midrule
V2A-MelSpec-S with audio pre-training \cite{large_scale_unsupervised_audio_pretraining_for_v2a} & 2.01          & 0.719         & 0.536          & 3.66              \\
\midrule

AV2A-MelSpec-S with audio pre-training & 2.02          & 0.719         & 0.539        & 3.40              \\
\hspace{3mm} + modality dropout & \textbf{2.05}          & \textbf{0.721}         & \textbf{0.543}        & 3.32              \\
\hspace{3mm} + modality dropout (GT audio)          & 2.02         & 0.717        & 0.536 & 3.58              \\
\bottomrule
\end{tabular}
\end{adjustbox}
\label{table:av2a_grid_allspeakers_seen_speakers}
\end{table}

\begin{table}[h]
\captionsetup{justification=centering}
\caption{Results on TCD-TIMIT (3 lipspeakers, seen)}
\begin{adjustbox}{width=\columnwidth}
\begin{tabular}{@{}lcccc@{}}
\toprule
\textbf{Method}                   & \hspace{1cm} & \textbf{PESQ$\uparrow$} & \textbf{STOI$\uparrow$} & \textbf{ESTOI$\uparrow$} \\ \midrule \midrule
\textbf{Raw waveform models} \\ \midrule \midrule
End-to-end WGAN (2022) \cite{end_to_end_video_to_speech_synthesis_using_gans} &                           & 1.40             & 0.538             & 0.357                               \\
\midrule
V2A-WaveGAN \cite{large_scale_unsupervised_audio_pretraining_for_v2a} & & 1.41             & 0.552             & 0.364                               \\
\midrule
AV2A-WaveGAN & & 1.42          & 0.550         & 0.383                      \\
\hspace{3mm} + modality dropout &          & 1.41         & 0.547        & 0.358              \\
\hspace{3mm} + modality dropout (GT audio)&          & 1.43         & \textbf{0.569}        & 0.404              \\

\midrule
V2A-WaveGAN with audio pre-training \cite{large_scale_unsupervised_audio_pretraining_for_v2a} & & 1.43             & 0.562             & 0.395                             \\
\midrule
AV2A-WaveGAN with audio pre-training & & 1.43          & 0.558         & 0.403                      \\
\hspace{3mm} + modality dropout &          & 1.42         & 0.564        & 0.407              \\
\hspace{3mm} + modality dropout (GT audio)& & \textbf{1.44}          & 0.566         & \textbf{0.411}                      \\ \midrule \midrule

\textbf{Acoustic features models} \\ \midrule \midrule
VCA-GAN \cite{lip_to_speech_synthesis_with_visual_context_attention_gan} &                           & \textbf{1.43}             & \textbf{0.595}             & \textbf{0.420}  \\
Lip2Wav \cite{lip2wav} &                            & 1.35             & 0.558             & 0.365                               \\

\midrule
V2A-MelSpec-VS \cite{large_scale_unsupervised_audio_pretraining_for_v2a}& & 1.35          & 0.492         & 0.296                        \\
\midrule
AV2A-MelSpec-VS &          & 1.34         & 0.482        & 0.295              \\
\hspace{3mm} + modality dropout &          & 1.33         & 0.476        & 0.294              \\
\hspace{3mm} + modality dropout (GT audio) & & 1.34          & 0.485         & 0.290                      \\

\midrule
V2A-MelSpec-VS with audio pre-training \cite{large_scale_unsupervised_audio_pretraining_for_v2a} & & 1.39          & 0.503         & 0.328                        \\
\midrule
AV2A-MelSpec-VS with audio pre-training & & 1.39          & 0.503         & 0.335              \\
\hspace{3mm} + modality dropout &         & 1.39         & 0.494        & 0.316              \\
\hspace{3mm} + modality dropout (GT audio)&          & 1.38         & 0.495        & 0.318              \\

\bottomrule
\end{tabular}
\end{adjustbox}
\label{table:av2a_tcd_timit_lipspeakers}
\end{table}

In experiments with GRID (33 speakers, seen), shown in Table \ref{table:av2a_grid_allspeakers_seen_speakers}, we observe that AV2A-WaveGAN with audio pre-training and modality dropout (GT audio) outperforms all other comparable methods across reconstruction metrics. Although its WER (2.65\%) is lower than the WER of its base V2A model (2.83\%), it is higher than AV2A-WaveGAN with modality dropout (GT audio) which is at 2.50\%. As with GRID (4 speakers, seen), we observe that all three training procedures of AV2A-WaveGAN produce improved reconstruction metrics compared to the corresponding base V2A-WaveGAN models, whereas the WER fluctuates more. AV2A-MelSpec-S with audio pre-training and modality dropout outperforms all other works across reconstruction metrics; however, SVTS-S achieves a lower WER.

With the TCD-TIMIT (3 lipspeakers, seen) split, AV2A-WaveGAN with audio pre-training and modality dropout (GT audio) outperforms other raw waveform models on PESQ and STOI, but AV2A-WaveGAN and modality dropout (GT audio) achieves the highest STOI. Note that both models achieve higher metrics than their corresponding base V2A model. However, we observe that most of our AV2A models in the mel spectrogram domain do not achieve higher reconstruction metrics than their base V2A model. Only AV2A-MelSpec-VS with audio pre-training achieves equal PESQ and STOI to its base V2A model, and a higher ESTOI. In addition, our models are outperformed on this dataset by VCA-GAN\cite{lip_to_speech_synthesis_with_visual_context_attention_gan}.

\subsection{Results on unseen speakers}

Tables \ref{table:av2a_grid_unseen_speakers} and \ref{table:av2a_lrw} show our results on datasets with unseen speaker splits. On GRID (33 speakers, unseen)  (Table \ref{table:av2a_grid_unseen_speakers}) AV2A-WaveGAN with modality dropout produces the highest PESQ and STOI, while the equivalent model with audio pre-training produces the highest ESTOI. However, the End-to-end WGAN in \cite{end_to_end_video_to_speech_synthesis_using_gans} reports a lower WER than our models. We note that AV2A-WaveGAN shows a reduction in WER compared to the base V2A-WaveGAN model. AV2A-MelSpec-S with modality dropout (GT audio) produces the highest reconstruction metrics among all comparable methods, while the equivalent model with audio pre-training shows the best WER.

\begin{table}[h]
\captionsetup{justification=centering}
\caption{Results on GRID (33 speakers, unseen)}
\begin{adjustbox}{width=\columnwidth}
\begin{tabular}{@{}lcccc@{}}
\toprule
\textbf{Method}                    & \textbf{PESQ$\uparrow$} & \textbf{STOI$\uparrow$} & \textbf{ESTOI$\uparrow$} & \textbf{WER (\%)$\downarrow$} \\ \midrule \midrule
\textbf{Raw waveform models} \\ \midrule \midrule
End-to-end WGAN (2018) \cite{video_driven_speech_reconstruction_using_gans}                           & 1.26             & 0.494             & 0.198              & 32.76                 \\
End-to-end WGAN (2022) \cite{end_to_end_video_to_speech_synthesis_using_gans}                            & 1.37             & 0.568             & 0.289              & \textbf{16.05}                 \\

\midrule
V2A-WaveGAN \cite{large_scale_unsupervised_audio_pretraining_for_v2a} & 1.43          & 0.589         & 0.316      & 19.88 \\
\midrule
AV2A-WaveGAN & 1.42          & 0.592         & 0.325        & 18.97              \\
\hspace{3mm} + modality dropout & \textbf{1.45}          & \textbf{0.598}         & 0.332        & 18.32              \\
\hspace{3mm} + modality dropout (GT audio)& 1.44          & 0.597         & 0.327        & 18.83              \\

\midrule
V2A-WaveGAN with audio pre-training \cite{large_scale_unsupervised_audio_pretraining_for_v2a} & 1.43          & 0.595         & 0.326          & 17.63              \\
\midrule
AV2A-WaveGAN with audio pre-training & 1.42          & 0.595         & 0.330        & 18.50             \\
\hspace{3mm} + modality dropout & 1.42          & 0.595         & 0.330        & 18.13              \\
\hspace{3mm} + modality dropout (GT audio)& 1.43          & 0.593         & \textbf{0.337}        & 17.70              \\

\midrule \midrule
\textbf{Acoustic features models} \\ \midrule \midrule
Vid2Voc \cite{vocoder_based_speech_synthesis}   & 1.26             & 0.541             & 0.227              & 38.15 \\
VCA-GAN \cite{lip_to_speech_synthesis_with_visual_context_attention_gan}                            & 1.39             & 0.570             & 0.283              & 24.52                 \\
Visual Voice Memory \cite{speech_reconstruction_visual_voice_memory}          & 1.33             & 0.531             & 0.271              & 26.11                 \\
SVTS-S \cite{svts}                             & 1.40             & 0.588             & 0.318              & 17.84 \\

\midrule
V2A-MelSpec-S \cite{large_scale_unsupervised_audio_pretraining_for_v2a} & 1.40          & 0.594         & 0.322          & 18.00              \\
\midrule
AV2A-MelSpec-S & 1.43          & 0.602         & 0.340        & 17.69              \\
\hspace{3mm} + modality dropout & 1.40          & 0.599         & 0.332        & 18.44              \\
\hspace{3mm} + modality dropout (GT audio)& \textbf{1.45}          & \textbf{0.604}         & \textbf{0.342}        & 18.04              \\

\midrule
V2A-MelSpec-S with audio pre-training \cite{large_scale_unsupervised_audio_pretraining_for_v2a} & 1.43          & 0.598         & 0.335          & 17.90              \\
\midrule
AV2A-MelSpec-S with audio pre-training & 1.42          & 0.596         & 0.333        & 18.39              \\
\hspace{3mm} + modality dropout & 1.44          & 0.599         & 0.334        & 18.23              \\
\hspace{3mm} + modality dropout (GT audio)& 1.44          & 0.595         & 0.337        & \textbf{17.74}              \\

\bottomrule
\end{tabular}
\end{adjustbox}
\label{table:av2a_grid_unseen_speakers}
\end{table}

\begin{table}[h]
\captionsetup{justification=centering}
\caption{Results on LRW}
\begin{adjustbox}{width=\columnwidth}
\begin{tabular}{@{}lcccc@{}}
\toprule
\textbf{Method}                    & \textbf{PESQ$\uparrow$} & \textbf{STOI$\uparrow$} & \textbf{ESTOI$\uparrow$} & \textbf{WER (\%)$\downarrow$} \\ \midrule \midrule
\textbf{Raw waveform models} \\ \midrule \midrule
End-to-end WGAN (2022) \cite{end_to_end_video_to_speech_synthesis_using_gans}                            & 1.33             & 0.552             & 0.331              & 42.38                 \\
\midrule
V2A-WaveGAN \cite{large_scale_unsupervised_audio_pretraining_for_v2a} & 1.46          & 0.623         & 0.445        & 29.79 \\
\midrule
AV2A-WaveGAN & 1.47          & 0.623         & 0.446        & 30.73              \\
\hspace{3mm} + modality dropout & 1.46          & 0.623         & 0.450        & 27.47              \\
\hspace{3mm} + modality dropout (GT audio)& 1.48          & 0.632         & 0.459        & \textbf{24.96}              \\

\midrule
V2A-WaveGAN with audio pre-training \cite{large_scale_unsupervised_audio_pretraining_for_v2a} & 1.47          & 0.630         & 0.443          & 29.88             \\
\midrule
AV2A-WaveGAN with audio pre-training & 1.47          & 0.630         & 0.447        & 31.07              \\
\hspace{3mm} + modality dropout & \textbf{1.48}          & 0.626         & 0.447        & 28.19              \\
\hspace{3mm} + modality dropout (GT audio)& \textbf{1.48}          & \textbf{0.637}         & \textbf{0.463}        & 26.37              \\

\midrule \midrule
\textbf{Acoustic features models} \\ \midrule \midrule
VCA-GAN \cite{lip_to_speech_synthesis_with_visual_context_attention_gan}                            & 1.34             & 0.565             & 0.364              & 37.07                 \\
Lip2Wav \cite{lip2wav}          & 1.20             & 0.543             & 0.344              & 34.20\footnotemark[2]                 \\
SVTS-M \cite{svts}                             & 1.46             & 0.649             & 0.482              & \textbf{12.90}                 \\

\midrule
V2A-MelSpec-M \cite{large_scale_unsupervised_audio_pretraining_for_v2a} & 1.48          & 0.649         & 0.484          & 14.96              \\
\midrule
AV2A-MelSpec-M & \textbf{1.49}          & \textbf{0.656}         & \textbf{0.494}        & 14.24              \\
\hspace{3mm} + modality dropout & 1.48          & 0.653         & 0.486        & 15.20              \\
\hspace{3mm} + modality dropout (GT audio)& 1.47          & 0.652         & 0.484        & 15.52              \\

\midrule
V2A-MelSpec-M with audio pre-training \cite{large_scale_unsupervised_audio_pretraining_for_v2a} & 1.46          & 0.646         & 0.476          & 18.70              \\
\midrule
AV2A-MelSpec-M with audio pre-training & 1.47          & 0.652         & 0.483        & 18.44              \\
\hspace{3mm} + modality dropout & 1.47          & 0.648         & 0.476        & 19.41              \\
\hspace{3mm} + modality dropout (GT audio)& 1.46          & 0.648         & 0.474        & 19.96              \\

\bottomrule
\end{tabular}
\end{adjustbox}
\vfill
\hspace{3mm} \\
$^2$Reported using Google speech-to-text API
\label{table:av2a_lrw}
\end{table}

Finally, Table \ref{table:av2a_lrw} shows the results on LRW. AV2A-WaveGAN with audio pre-training and modality dropout (GT audio) reports the best reconstruction metrics among raw waveform models, but the equivalent model without audio pre-training show a lower WER. AV2A-MelSpec-M also outperforms comparable works across reconstruction metrics, but SVTS-M reports a lower WER. 

\section{Conclusion}
In this work we introduced a video-to-speech synthesis framework that uses video and audio inputs during both training and inference. This is accomplished in a two step-process; firstly, by using a pre-trained video-to-speech model to synthesize the missing audio and secondly, by using the video and the synthesized audio samples as inputs to a new, audio-visual-to-audio (AV2A) model. We proposed a simple method to obtain such a model, by appending an audio encoder to an existing video-to-speech model. We conduct experiments in both the raw waveform and mel spectrogram domains, and introduce two variants of modality dropout during training. Our experiments demonstrate the effectiveness of our approach, as measured by objective metrics.

In future work it would be interesting to extend this framework to other speech synthesis tasks, such as audio-visual speech enhancement, separation and inpainting (i.e., by using the synthesized outputs of such a model in the inputs of a second model). Inspired by the literature on speech enhancement\cite{an_overview_of_deep_learning_based_av_speech_enhancement_and_separation}, another beneficial research direction would be to experiment with different training objectives in the AV2A model, such as mask approximation and indirect mapping. Furthermore, one may also investigate mixing models designed for different tasks, such as using a V2A model to synthesize the audio and then conducting AV2A on an off-the-shelf audio-visual speech enhancement model.

\bibliographystyle{IEEEtran}
\bibliography{references}

\end{document}